\documentclass[12pt,preprint]{aastex}

\usepackage{graphicx,lscape,subfigure}
\usepackage{latexsym}
\usepackage{epsfig}

\begin{document}

\title{The Early-time Optical Properties of Gamma-Ray Burst Afterglows}

\author{A.~Melandri\altaffilmark{1}, C.G.~Mundell\altaffilmark{1}, S.~Kobayashi\altaffilmark{1}, C.~Guidorzi\altaffilmark{2,3,1}, A.~Gomboc\altaffilmark{4}, I.A.~Steele\altaffilmark{1}, R.J.~Smith\altaffilmark{1}, D.~Bersier\altaffilmark{1}, C.J.~Mottram\altaffilmark{1}, D.~Carter\altaffilmark{1}, M.F.~Bode\altaffilmark{1}, P.T.~O'Brien\altaffilmark{5}, N.R.~Tanvir\altaffilmark{5}, E.~Rol\altaffilmark{5}, R.~Chapman\altaffilmark{6}}
\email{axm@astro.livjm.ac.uk}

\altaffiltext{1}{Astrophysics Research Institute, Liverpool John Moores University, Twelve Quays House, Egerton Wharf, Birkenhead, CH41 1LD, UK}
\altaffiltext{2}{Universit\`a di Milano Bicocca, Dipartimento di Fisica, piazza della Scienze 3, I-20126 Milano, Italy}
\altaffiltext{3}{INAF Osservatorio Astronomico di Brera, via Bianchi 46, 23807 Merate (LC), Italy}
\altaffiltext{4}{FMF, University of Ljubljana, Jadranska 19, 1000 Ljubljana, Slovenia}
\altaffiltext{5}{Department of Physics and Astronomy, University of Leicester, University Road, Leicester LE1 7RH, UK}
\altaffiltext{6}{Centre for Astrophysics Research, University of Hertfordshire, College Lane, Hatfield AL10 9AB, UK}

\label{firstpage}

\begin{abstract}

We present a multiwavelength analysis of 63 Gamma-Ray Bursts observed
with the world's three largest robotic optical telescopes, the
Liverpool and Faulkes Telescopes (North and South). Optical emission
was detected for 24 GRBs with brightnesses ranging from $R$ = 10 to 22
mag in the first 10 minutes after the burst. By comparing optical and
X-ray light curves from t = 100 to $\sim 10^6$ seconds, we introduce
four main classes, defined by the presence or absence of temporal
breaks at optical and/or X-ray wavelengths. While 15/24 GRBs can be
modelled with the forward-shock model, explanation of the remaining
nine is very challenging in the standard framework even with the
introduction of energy injection or an ambient density gradient. Early
X-ray afterglows, even segments of light curves described by a
power-law, may be due to additional emission from the central
engine. 39 GRBs in our sample were not detected and have deep upper
limits ($R$ $<$ 22 mag) at early time. Of these, only ten were
identified by other facilities, primarily at near infrared
wavelengths, resulting in a dark burst fraction of
$\sim$50\%. Additional emission in the early time X-ray afterglow due
to late-time central engine activity may also explain some dark bursts
by making the bursts brighter than expected in the X-ray band compared
to the optical band.

\end{abstract}


\section{Introduction}

Gamma Ray Bursts (GRBs) are brief, intense flashes of high energy
gamma rays originating at cosmological distances and often associated
with subsequent radiation emitted at longer wavelengths from X-ray to
radio waves on times scales of minutes to days after the initial gamma
ray burst.  In the standard model a typical long duration GRB is
thought to be formed by the explosion of a compact source that
generates an expanding relativistic fireball (Rees $\&$
M{\'e}sz{\'a}ros 1992). If the central engine remains active for some
time several expanding shells with different speeds (different Lorentz
factors, $\Gamma$) can be produced. The collisions between these
shells power the $\gamma$-ray prompt emission itself (internal
shocks), while the interactions of the relativistic flow with the
surrounding medium (external forward shocks) generate the so-called
afterglow emission that dominates at longer wavelengths and is more
long-lived than the prompt emission (M{\'e}sz{\'a}ros 2002, Piran
1999). Assuming that the shock-accelerated electrons producing the
radiation have a power-law spectral energy distribution, the afterglow
synchrotron emission is expected to exhibit a standard form of
spectrum, with two characteristic break frequencies: the typical
synchrotron frequency $\nu_{\rm m}$ and the cooling frequency
$\nu_{\rm c}$ (Sari, Piran $\&$ Narayan 1998). When the forward shock
is formed, a reverse shock that propagates backwards into the ejecta
is also generated. The brightness of the reverse shock emission decays
very rapidly compared to the decay of the forward component. It is
predicted that at early time the reverse shock can produce extremely
bright optical flashes while at late time the optical flux is
completely dominated by the forward shock emission (M{\'e}sz{\'a}ros
$\&$ Rees 1997, Sari $\&$ Piran 1999). In reality, the resultant light
curve is a complex, time-dependent mixture of these components and
unravelling them provides important insight into the physics and
energetics of the explosion.

The study of pre-{\it Swift} bursts allowed the understanding of late
time multiband properties of the afterglows confirming many
predictions of the fireball model. Well-sampled pre-{\it Swift}
optical light curves were mostly obtained at late times, typically
after few hours from the burst event, and exhibited relatively smooth
light curves with simple power law decays, showing breaks at late time
(hints of a jet evolution) and making clear a strong connection with
supernova emission and, thus, the death of massive stars (e.g. Woosley
$\&$ Bloom 2006, Malesani et al. 2004, Stanek et al. 2003). The advent
of the {\it Swift} satellite (Gehrels et al. 2004) has opened up a new
observational window at early times, revealing more complex light
curve behaviour than previously known. It is now accepted that the
X-ray temporal decay of GRBs observed by {\it Swift} is well described
by a canonical light curve (Nousek et al. 2006; Zhang et al. 2006);
combining the $\gamma$-ray \citep[Burst Alert Telescope, BAT,][]{bart}
and X-ray (X-Ray Telescope, XRT, Burrows et al. 2005) data the initial
X-ray emission (rapid fall-off for the first hundred seconds) is
consistent with the tail of the $\gamma$-ray emission (Tagliaferri et
al. 2005, O'Brien et al. 2006) and can be modelled by two components
that have exactly the same functional form (Willingale et
al. 2007). These functions are completely empirical and do not do not
provide a physical explanation for the X-ray flares seen in many
bursts, but the majority of {\it Swift} bursts seem to follow this
behaviour.

It was clear that rapid response to obtain early-time optical
observations with ground based robotic telescopes was required to pin
down the open issue of the emission mechanism for the GRB itself and
its afterglow. However, the small number of prompt optical
observations simultaneous to the GRB $\gamma$-ray emission do not yet
allow a firm conclusion to be drawn about the GRB emission models
(Yost et al. 2007). The statistics of these events remain small due to
the fact that the GRBs detected by {\it Swift} satellite are fainter
and located at higher redshift ($<z> = 2.7$, Le $\&$ Dermer 2007,
Jakobsson et al. 2006) than those detected by previous missions. Large
2-m robotic telescopes, such as the Liverpool and Faulkes (North and
South) Telescopes, responding within 1-3 minutes of the burst offer a
unique tool for probing early-time light curves over a wide range of
brightnesses, allowing the extension of the analysis of GRB properties
to unprecedented depth and time coverage.

Based on the theoretical predictions of the forward and reverse shock
emission theories, optical light curves at early times should show
different shapes depending on the relative contribution of the two
components (Kobayashi $\&$ Zhang 2003, Zhang, Kobayashi $\&$
M{\'e}sz{\'a}ros 2003); possible light curve shapes are illustrated in
Figure~\ref{figc1}. In particular, if the optical observations start
when the reverse shock component still dominates the optical flux, the
shape of the optical light curve will appear as case 1 or case 2 in
Figure~\ref{figc1}. In the first case the light curve will show a
transition from steep to shallow power law decay index. In the second
case it will show a re-brightening, but if observations do not start
early enough the first steep segment (thick dashed line on
Figure~\ref{figc1}) of the light curve should be missing and it will
be visible only in the rise of the forward component. The observed
peak in case 2 will correspond to the passage of the $\nu_{\rm m}$ in
the observing band.  If there is energy injection to the forward shock
emission then the light curve should appear as case 3 in
Figure~\ref{figc1}.  This behaviour can be explained by long-lived
central engine activity or large dispersion in the distribution of
Lorentz factors (Zhang et al. 2006) or related to the time when the
energy is transferred from the fireball to the ambient medium
(Kobayashi $\&$ Zhang 2007).

Within this theoretical framework, we present the analysis of a sample
of 63 GRBs observed with the network of three 2-m telescopes
\emph{RoboNet-1.0}\footnote[1]{\it
http://www.astro.livjm.ac.uk/RoboNet/} (Gomboc at al. 2006), formed by
the Liverpool Telescope (LT, La Palma, Canary Islands), the Faulkes
Telescope North (FTN, Haleakala, Hawaii) and the Faulkes Telescope
South (FTS, Siding Spring, Australia). For those GRBs with detected
optical counterparst, we discuss their light curve properties, compare
optical and X-ray data and analyse the intrinsic rest frame properties
of those bursts with known spectroscopic redshift. An analysis of the
bursts for which no afterglow was detected is also presented and
discussed within the standard fireball model.

Throughout we use the following conventions : the power law flux is
given as $F(\nu,t) \propto t^{-\alpha} \nu^{-\beta}$, where $\alpha$
is the temporal decay index and $\beta$ is the spectral slope; we
assume a standard cosmology with $H_{0} = 70$~km~s$^{-1}$~Mpc$^{-1}$,
$\Omega_{m} = 0.3$ and $\Omega_{\Lambda} = 0.7$; all errors and
uncertainties are quoted at the $1\sigma$ confidence level.

\section{Observations and Analysis}

All three telescopes enable rapid response (the typical mean reaction
time is $<t> \sim 2.5$ minutes after the trigger) and deep
observations ($R \sim$ 21 at $t \sim 5$ minutes after the trigger) to
GRB alerts, which are crucial in the case of faint or optically dark
bursts. Each telescope operates in a fully robotic mode, responding
automatically to a GRB satellite alert by immediately over-riding the
current observing programme, then obtaining, analysing and
interpreting optical images of the GRB field using the specially
designed, sophisticated pipeline (LT-TRAP); subsequent robotic
followup observations are then optimised and driven by the
automatically derived properties of the afterglow (see Guidorzi et
al. 2006). Up to September 2007 the network robotically reacted to 63
GRBs with 24 optical afterglow detections and 39 upper limits (Figure
\ref{figall}).

In this paper, the optical photometry of each optical afterglow was
performed using Starlink/GAIA photometry tools. Each field of view was
calibrated with the best data available: 1) standards stars observed
during the night, if the night was photometric; 2) pre-burst
calibration fields (both for SDSS or Bessell filters) reported in GCN
Circulars; 3) differential photometry with catalogued stars in the
field of view (USNOB1.0 catalogue). This procedure has been followed
for all the sample data. All the calibrated magnitudes in the SDSS-r'
filter (for all the bursts observed with the LT telescope) were then
transformed to the $R_{\rm C}$ band using the filter transformations
given by Smith et al. (2002); the colour term $R-r'$ was then derived
for the selected stars used to calibrate the field and finally applied
to the estimated magnitude. Data calibration is also discussed in
Guidorzi et al. (2005b, 2007), Monfardini et al. (2006) and Mundell et
al. (2007b).

\section{Results}

\subsection{Optical Detections}

The observed optical light curves of the detected afterglows in our
sample are shown in Figure~\ref{figall} (left panel). In the right
panel of the same figure all the deep upper limits for the undetected
afterglows are shown, using the same scale on the y-axis to emphasize
the large range in brightness and time covered by our observations.

The individual light curves of the detected afterglows are then shown
in Figure~\ref{figxopt}, which shows $R$ band observations together
with the corresponding {\it Swift} X-ray light curves. Temporal and
spectral properties are summarised in Table 1. We give both temporal
and spectral information for each burst, when available. We fitted the
data either with a single or with a broken power law; the values
reported in the table are the best fitting results. The values
reported for $\beta_{\rm O}$, $\beta_{\rm X}$ and $\beta_{\rm OX}$ are
retrieved from the literature. In the last two columns of the table we
report the redshift ($z$) and the derived isotropic gamma-ray energy
($E_{\gamma, {\rm iso}}$) of the burst. For all the bursts detected by
the {\it Swift} satellite the value of $E_{\gamma, {\rm iso}}$
reported in the table is taken from Butler et al. (2007). For the
no-{\it Swift} bursts we calculated $E_{\gamma, {\rm iso}}$ assuming a
standard cosmology as reported at the end of Section 1, with the
following formula: $E_{\gamma,iso} = (4 \pi D_{L}^{2}~{\it f}) / (1 +
z) $, where $D_{L}$ is the luminosity distance and {\it f} is the
gamma ray fluence of the burst.

\begin{figure*} \centering 
\includegraphics[height=7.5cm,width=14cm]{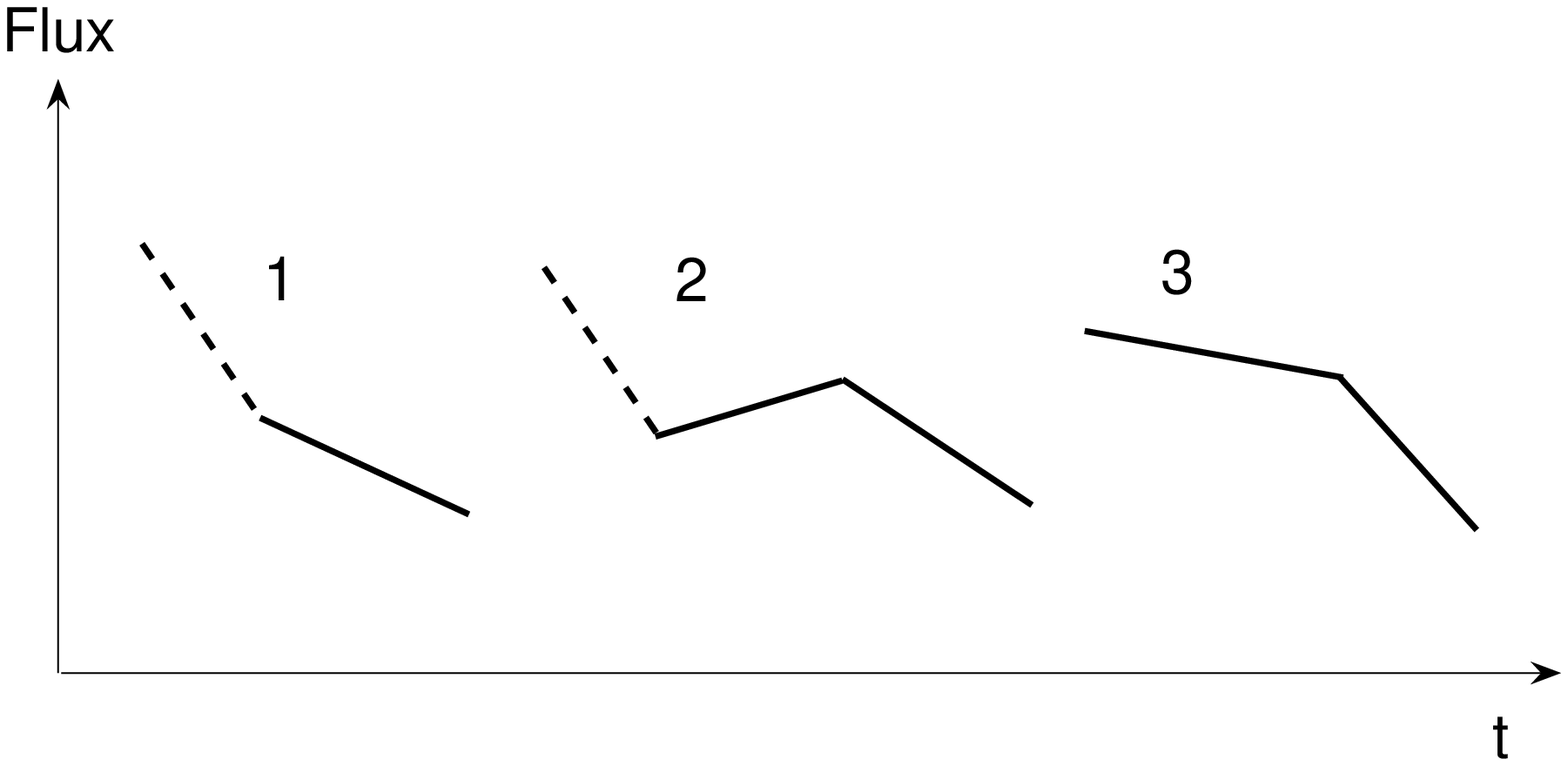}
\caption{Schematic illustrating possible shapes of the optical light 
curves at early times as a result of the contribution of reverse and
forward shock emissions (case 1 and 2) or due to energy injection
(case 3). The thick dashed line for case 1 and 2 represent the reverse
shock contribution at early times, that can be missing if the
observations do not start early enough.}
\label{figc1} 
\end{figure*}

\begin{figure*} \centering 
\includegraphics[height=7cm,width=7cm]{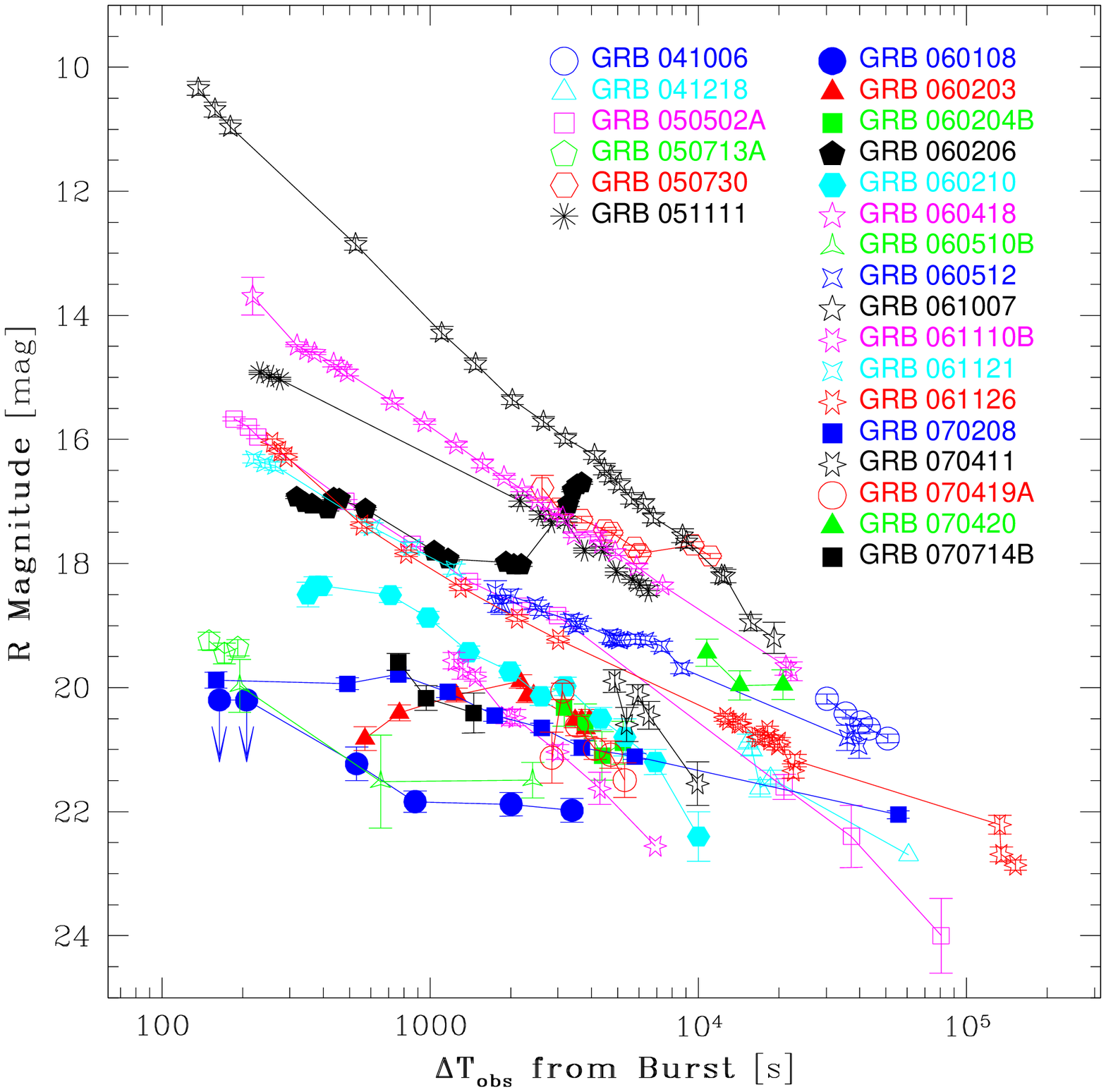}
\includegraphics[height=7cm,width=7cm]{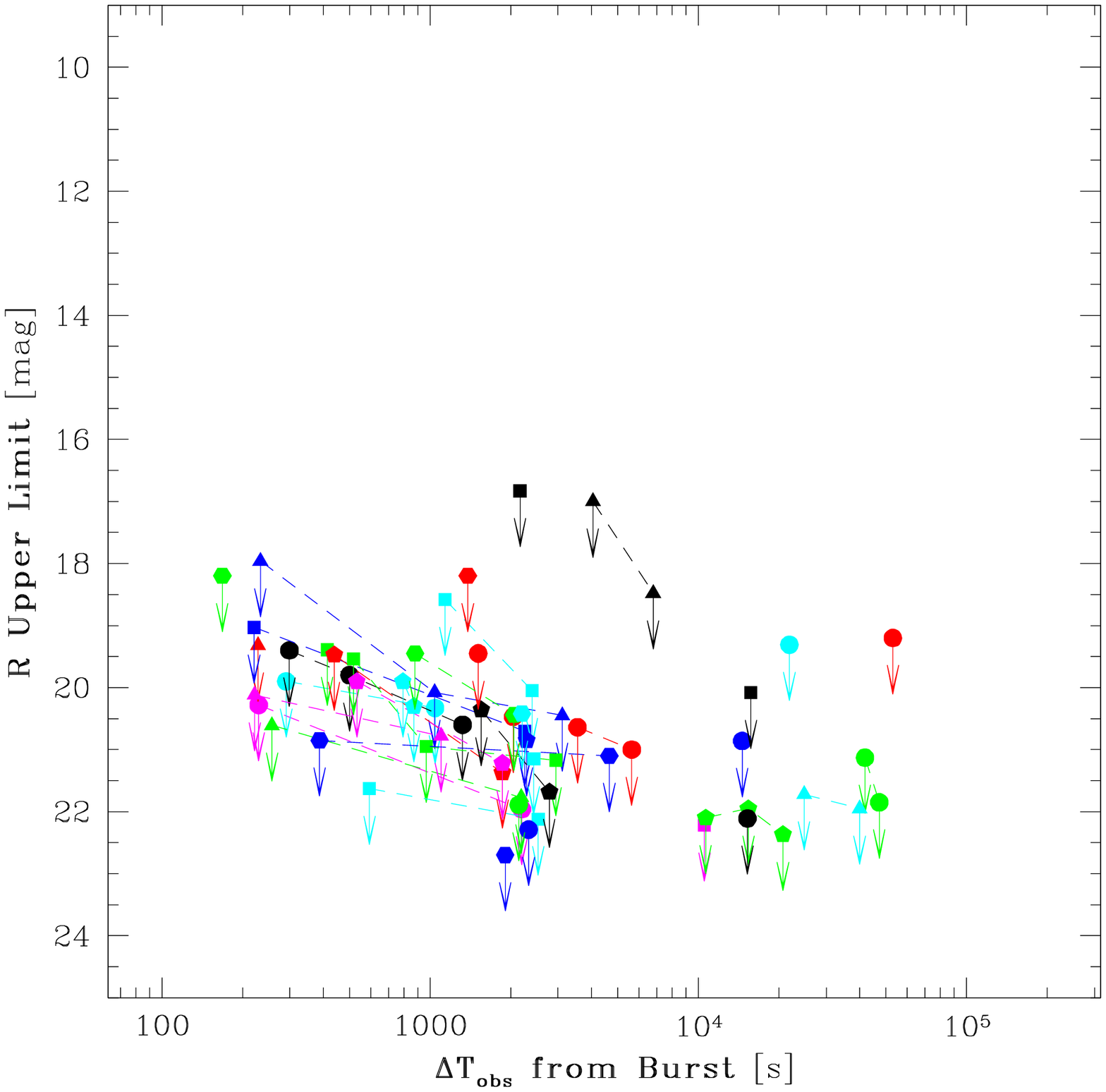}
\caption{Left panel: observed light curves in the $R$ filter of 
all the detected afterglows in our sample. GRB~060927 is not included
as we detected this burst only in the $i'$ band (see Ruiz-Velasco et
al. 2007). Right panel : optical upper limits in the $R$ band of the
remaining GRBs observed with, but not detected by the Liverpool and
Faulkes telescopes. Connected symbols refer to different observations
for the same burst when additional late time observations were
available.}
\label{figall} 
\end{figure*}

Some of the sample bursts have been discussed in previous dedicated
papers, but are presented here for completeness. For each burst in our
sample we summarise its key properties, together with references to
more detailed work where relevant:

\begin{itemize}

\item GRB~041006 : this HETE burst was observed with Chandra between
16.8 and 42.5 hours after the burst event, showing a single temporal
power law decay, whereas in the optical a break in the light curve at
early time was clearly visible. At late time ($>10^{6}$ s) the
contribution of the underlying supernova emerged (Stanek et al. 2005,
Soderberg et al. 2006). The spectral indices in the two bands are
consistent, with a slope $\beta_{\rm OX} \sim 0.7$ (Butler at
al. 2005; Garnavich et al. 2004). Chandra X-ray data are not shown in
our plot; the results of X-ray analysis are taken from Butler et
al. 2005.

\item GRB~041218 : for this long INTEGRAL burst there is only a late XRT
observation. We do not have information for the X-ray temporal decay
but in the optical we see a steepening of the decay after $\sim 0.10$
days. No spectral information is available for this burst.

\item GRB~050502A : for this long INTEGRAL burst the XRT provided only
an upper limit for the X-ray flux at $\sim 1.3$ days. In the optical
band the light curve can be described by a single power law with a
hint of a bump at 0.02 days. In the X-ray band, temporal and spectral
parameters are consistent with similar behaviour to that of the
optical (see Guidorzi et al. 2005b for the detailed analysis of this
burst).

\item GRB~050713A : XMM observations performed between 5.8 and 13.9 
hours after the burst, show a break in the X-ray light curve at $\sim
2 \times 10^{4}$ s. That break is marginally detected in the X-ray
data acquired by {\it Swift}, while the optical light curve is well
described by a shallow power law at all times (Guetta et al.,
2006). XMM data are not reported in our plot; an accurate spectral
analysis has been done by Morris et al. (2007).

\item GRB~050730 : both X-ray and optical light curves of this 
{\it Swift} burst show a steepening of the initial temporal power law
decay at 0.1 days, more pronounced in the X-ray band. The derived
spectral slopes from X-ray and optical data are statistically in
agreement (Pandey et al. 2006).

\item GRB~051111 : the optical light curve is well fitted by a broken 
power law with an early break around 12 minutes, while in the X-ray no
early data were acquired and no break is visible. A comprehensive
multi-wavelength temporal and spectral study of this burst has been
done by Guidorzi et al. (2007), Yost et al. (2007) and Butler at
al. (2006).

\item GRB~060108 : this burst has a faint, but relatively blue,
optical afterglow that was identified by deep rapid FTN observation,
without which it would have been classified as dark burst ($\beta_{\rm
OX} \sim 0.5$ and no afterglow was detected by the {\it Swift} UVOT
within the first 300 s). It shows a canonical X-ray light curve while
in the optical band the flux may show a similar behaviour but, due to
its optical faintness the resultant sparse sampling with co-added
images provides a light curve that is consistent wit a single power
law. The darkness of this burst can be explained by a combination of
an intrinsically optical faintness, an hard optical to X-ray spectrum
and a moderate extinction in the host galaxy (Oates et al. 2006).

\item GRB~060203 : after an initial rise, the optical light curve 
follows a shallow power law decay (typical example of the case 2
reported in Figure~\ref{figc1}). Due to observing constraints, X-ray
observations started only $10^{3}$~s after the trigger and the light
curve shows evidence of a power law decay roughly consistent with the
optical one. The spectral fit gives a value of $\beta_{\rm X} \sim
1.1$ and shows a value of N$_{\rm H}$ in excess of the galactic one
(Morris et al. 2006a).

\item GRB~060204B : the early X-ray light curve is dominated by 
flaring activity and only after $2\times10^{3}$~s does the decay
appear to be a single power law (Falcone et al. 2006a). In the optical
the afterglow is faint and the light curve is a simple power law,
shallower than the X-ray decay.

\item GRB~060206 : the detailed analysis of the optical light curve
showed dramatic energy injection at early time, with at least 3
episodes, with the largest one also visible in the X-ray light
curve. At late times a break is evident in the optical but not in the
X-ray band; the chromatic nature of the break is not consistent with
the possible jet-break interpretation and can be ascribed to a change
in the circumburst density profile (Monfardini et al. 2006). In a
recent work Curran et al. (2007) found that the data are also
consistent with an achromatic break, even if the break in the X-ray is
less pronounced. However, the complete X-ray data set acquired with
XRT remains consistent with a single unbroken power law decay up to
$10^{6}$ seconds (Morris et al. 2006b, Burrows $\&$ Racusin 2007) ,
and the first explanation for the break seems to be the more
plausible.

\item GRB~060210 : the optical light curve exhibits a power law decay 
after an initial flat phase ($\alpha = 0.09 \pm 0.05$, case 3 in
Figure~\ref{figc1}). There is marginal evidence for a break at late
times ($t_{{\rm O},break}$ $\sim 0.1 $ days) but there are not enough
optical data to support that. Instead a break is visible in the X-ray
light curve, but the late time break is not simultaneous ($t_{{\rm
X},break} \sim 0.3$ days). We can not exclude the possible achromatic
nature of that break due to the very complex behaviour of X-ray light
curve.

\item GRB~060418 : in the optical and infrared bands (see Molinari et
al., 2007 for the infrared analysis) an initial rise is visible,
followed by a straight power law decay (case 2 in
Figure~\ref{figc1}). Fitting the early X-ray data it is possible to
see a change in the slope, but, accounting for the presence of a large
flare, the data are also consistent with a single power law
decay. There is no evidence of temporal break up to $10^6$~s in both
optical and X-ray bands. The spectrum of the underlying X-ray
afterglow can be described by a simple absorbed power law with
$\beta_{\rm X} = 1.04\pm0.13$ (Falcone et al. 2006b). The low degree
of polarization of the optical light at early time (at $\sim 200$ s
after the event) ruled out the presence of a large scale ordered
magnetic field in the emitting region (Mundell et al. 2007a).

\item GRB~060510B : the temporal behaviour in the X-ray seems to be 
the canonical steep-shallow-steep decay with superimposed flares, even
if the light curve at late times is poorly sampled. Looking at the
flat light curve at early times, possibly the prompt gamma-ray
component is detected in XRT (T$_{90} \sim 280$ s). Due to the high
redshift, the optical flux is suppressed by the Ly$\alpha$ absorption
and our early optical data are consistent with a single power law
decay. No optical observations at late time were available to confirm
the possible achromatic nature of the break observed in the X-ray
band. The early time X-ray spectrum is well fitted with an absorbed
power law with $\beta_{\rm X} \sim 0.5$ (Perri et al. 2006). This is
the value of $\beta_{\rm X}$ reported in Table1 because
estimated at a time consistent with the time of our optical
observations. At late times, taking into account the intrinsic
absorption at the redshift of the burst, the spectrum is well fitted
by an absorbed power law with $\beta_{\rm X} \sim 1.5$ (Campana et
al., 2006).

\item GRB~060512 : optical data are consistent with a single power law.
In the X-ray a clear flare is visible at early times and the decay
after $10^{3}$ s is a simple power law. At late time, the X-ray
spectrum has a slope $\beta_{\rm X} \sim 0.9$ (Godet et al. 2006).

\item GRB~060927 : due to the high redshift ($z=5.467$), the optical 
light curve is highly affected by Ly$\alpha$ suppression, particularly
in the $R$ filter. The light curve in the $I$ band shows evidence of a
possible extra component at early times, but after 500 s the decay is
a single power law ($\alpha_{\rm I} \sim 1.2$). In the X-ray a change
of slope is clearly visible and the spectrum is well modelled with an
absorbed power law with $\beta_{\rm X} \sim 0.9$ (Ruiz-Velasco et al.,
2007).

\item GRB~061007 : the optical light curve exhibits an early peak
followed by an unprecedented straight power law decay up to (and
likely beyond) $10^6$~s after the burst (a good example of case 2 in
Figure~\ref{figc1}), perfectly mirrored in the X-ray band. The peak at
early times can be explained in the context of the fireball model: no
optical flash is seen because the typical frequency of the reverse
shock emission lies in the radio band at early time and the optical
afterglow is dominated by forward shock emission (Mundell et
al. 2007b).  The broad band optical to $\gamma$-ray spectral energy
distribution is well described by an absorbed power law with $\beta
\sim 1.0$ (Mundell et al. 2007b, Schady et al. 2007).

\item GRB~061110B : this burst is intrinsically faint and displays a 
simple power law decay both in the optical and in the X-ray bands. It
showed a typical GRB afterglow spectrum with $\beta_{\rm X} \sim 1.0$
(Grupe et al. 2006).

\item GRB~061121 : this is a perfect example of canonical light curve 
in the X-ray band not replicated in the optical. The light curve
follows a simple power law decay even though the observations began
when the initial steep X-ray phase was still ongoing (Page et al.,
2007).

\item GRB~061126 : the optical light curve shows a steep to shallow 
transition at about 13 minutes after the trigger. The early, steep
component can be interpreted as due to the reverse shock while the
later slowly fading component is coming from the forward shock (clear
example of case 1 in Figure~\ref{figc1}). X-ray observations started
after the transition in optical, and show the X-ray afterglow decaying
with $\alpha_{{\rm X},2} > \alpha_{{\rm O},2}$ (faster than the
optical afterglow) to the end of observations at $10^{6}$ s (for more
details see Gomboc et al. 2008).

\item GRB~070208 : after an initial rising phase, the X-ray light curve 
shows a power law decay. The optical data cover the same time interval
as the X-ray data but no rising phase is detected and the optical
light curve is well described by a simple power law, after an initial
flat phase (case 3 in Figure~\ref{figc1}). The optical decay index
after the flat phase remains shallower than the decay in the X-ray
band.

\item GRB~070411 : in this case the temporal decay in the two bands is
very similar, but in the optical band an initial rising phase is
detected that is not visible in the X-ray band, probably due to the
poor sampling at early times in that band. The optical light curve
seems to be another example of case 2 reported in Figure~\ref{figc1}
but the shape is not smooth and clear, with significant scatter
(possibly variability) around the power law decay.

\item GRB~070419A : different observations in the optical band show
that at early time the afterglow brightened before starting a shallow
decay phase that lasts up to $10^{6}$~s. Although the X-ray temporal
decay at late time ($\alpha_{{\rm X},2}$ = $0.64\pm0.10$) agrees very
well with the optical decay ($\alpha_{{\rm O},2}$ = $0.58\pm0.04$), at
early times the X-ray light curve shows a rapid decay, with no hints
of any flare activity. The shape observed in the optical band does not
fit the three cases reported in Figure~\ref{figc1} and is probably the
result of an episode of energy injection.

\item GRB~070420 : another example of canonical light curve in the X-ray 
band not replicated in the optical. However, in this case the optical
light curve is less well sampled than GRB~061121, so it is more
difficult to constrain the optical behaviour in the entire time
interval covered by XRT observations.
 
\item GRB~070714B : the X-ray data show a steep-shallow-steep decay, 
not well constrained at late times and with possible small flaring
activity. The optical counterpart is very faint but appears to show a
shallower power law decay compared with the X-ray decay at late times.

\end{itemize}

\subsection{Optical Upper-Limits}

In Table~2 we report the optical upper limits for all the bursts
observed with the Liverpool and the Faulkes telescopes for which we
did not detect any afterglow candidate. For each GRB we specified when
the XRT position for the X-ray afterglow was found. The duration, the
BAT fluence $\it f$ (15-150 keV), the XRT early flux F$_{\rm X}$
(0.3-10 keV), the time of X-rays observations ($\Delta~t_{\rm X}$),
the temporal decay ($\alpha_{\rm X}$) and the spectral slope
($\beta_{\rm X}$) in the X-ray band, are the values reported in the
GCNs or taken from the {\it Swift} general table. $R^{\rm u.l.}_{\rm
start}$ are the values of our optical upper limits at $\Delta~t_{\rm
start}$ minutes from the trigger, where $\Delta~t_{\rm start}$ is the
starting time of our observations. $R^{\rm u.l.}_{\rm mean}$ are the
values of our optical upper limits at $\Delta~t_{\rm mean}$ minutes
from the trigger (mean time) for the coadded frames with a total
integration time of $T_{exp}$ minutes. The columns OT and $A_{R}$ show
when an optical (O) or infrared (IR) afterglow for that burst had been
detected by other facilities and the extinction for the $R$ band in
the direction of the burst. The last two columns on the table are the
value of the X-ray temporal decay inferred from our fit of the XRT
light curves ($\alpha_{\rm X}^{\rm (fit)}$) and the estimate of the
X-ray flux (F$_{\rm X}$) at the time of our optical upper limit
$\Delta~t_{\rm mean}$.

X-ray light curves are given in Figure~\ref{figul2} together with our
optical upper limit. The best fit for the X-ray light curves is
shown. The temporal decay indices in the X-ray band ($\alpha_{\rm
X}^{\rm (fit)}$) reported in Table~2 refer to the segment of the light
curve contemporaneous with our optical limit. Only two of the bursts
listed in the table were detected by UVOT in the optical bands
(GRB~070721A and GRB~070721B); we did not detect the optical
counterpart for these two bursts because our observations were
performed at late time due to observational constraints.

Of the 39 non-detections, 10 were detected by other facilities,
primarily at infrared wavelengths or using larger aperture optical
telescopes; the other 29 remain as non-detections. Details are
summarised below:

\begin{itemize}

\item GRB~050124 : an infrared candidate was detected and confirmed 
by two Keck observations performed in the $K_{s}$ band about 24.6 and
47.8 hours after the burst ($\Delta K_{s} \sim 0.5$, Berger et
al. 2005a, Berger et al. 2005b). No afterglow was detected in the
optical bands even by the UVOT telescope ($\sim 3$ hours after the
trigger, Lin et al. 2005, Hunsberger et al. 2005). Our observations
were performed manually in the $R$ band only after 14.7 hours for
observational constraints and no optical counterpart was detected.

\item GRB~050716 : our observations in the optical band began
 3.8 minutes after the burst with the FTN telescope but no optical
 candidate was found down to a limit of R$\sim$20 mag (Guidorzi et
 al. 2005a). A potential infrared counterpart ($J-K \sim 2.5$) was
 found in UKIRT observations just outside the XRT error circle (Tanvir
 et al. 2005). At the position of that candidate we did not clearly
 detect any source in the $R$ and $I$ bands but we found an excess
 flux which suggests that the afterglow is probably reddened rather
 than at very high redshift. A broad-band analysis found that $z \sim$
 2 is a good estimate for the redshift of this event and that a host
 galaxy extinction of A$_{\rm V} \sim 2.0$ can account for the
 relatively faint optical/infrared afterglow observed (Rol et
 al. 2007b).

\item GRB~060116 : also for this burst an infrared candidate was 
detected in UKIRT observations (Kocevski et al. 2006a, Kocevski et
al. 2006b). The afterglow of that burst was detected with an
unfiltered magnitude $\sim$20 (Swan et al. 2006) and the very red
colours of this afterglow ($J-K =2.5$, $I-J > 2.9$, Malesani et
al. 2006a) suggest the source to be a highly dust extincted (A$_{\rm
V} > 2.5$) but not at high redshift (Tanvir at al. 2006a).

\item GRB~060121: the SXC error circle for this HETE-II short burst 
was partially covered by our first LT observation starting 0.83 hours
after the burst. Only in the second observation starting 2.44 hours
after the event the XRT error circle was entirely inside our field of
view. The detected optical/infrared afterglow (Levan et al. 2006a,
Malesani et al. 2006b, Hearty et al. 2006a, Hearty et al. 2006b) is
not seen in our images down to a limiting magnitude of $R \sim 22$ at
$\sim 3$ hours after the burst. Subsequent HST observations revealed
the presence of a very faint (both in optical and infrared) red
galaxy, probably an edge-on disk, close to the position of the
afterglow (Levan et al. 2006). In this case this would favour a
higher redshift for this burst than has been measured for most short
bursts to date.

\item GRB~060319 : an infrared candidate was found in WHT observations 
($K = 19.0 \pm 0.3$, Tanvir et al. 2006b) but no claims about
variability have been made. Our observations set an upper limit of $R
\sim 21$ mag after 10 minutes of the burst and no other optical
observation detected any possible counterpart down to $R \sim 23$ mag
(D'Avanzo et al. 2006a, Lipunov et al. 2006).

\item GRB~060602A : a very faint possible optical candidate was detected 
about 15 minutes after the burst ($R \sim 22.5 \pm 0.3$, Jensen et
al. 2006). The transient nature of the source was difficult to assess
but was not visible in the SDSS pre-burst image. Our observations were
affected by the bright moon and only an upper limit of $R \sim 16.8$
mag was set about 36 minutes after the burst.

\item GRB~060923A : an infrared afterglow was detected in the $K$-band 
but undetected in $I$ and $J$ bands (Tanvir et al. 2008, Fox et
al. 2006a, Fox 2006), suggesting again a highly extinguished or high
redshift afterglow for this burst. The subsequent detection of a faint
host galaxy ($R\sim25.5$, Tanvir et al. 2008) set an upper limit for
the redshift of this burst of $z \sim 5$, leaving the extinction as
the most likely cause of the extremely red colours measured at early
times.

\item GRB~060923C : the afterglow of that burst was detected and 
confirmed in the infrared bands (Covino et al. 2006, Fox et al. 2006b,
D'Avanzo et al. 2006b). No indication of any optical identification
means again that the possible explanation is a high redshift or high
extinction origin for that burst.

\item GRB~061006 : a source was found to vary with $\Delta I \sim 0.5$ 
mag between 0.6 and 1.6 days after the burst in the $I$-band (Malesani
et al. 2006c, Malesani et al. 2006d). This source was identified with
the afterglow of this short-hard burst. However the inferred power-law
decay slope was quite shallow ($\sim 0.5$) and in the second
observation the source was extended so the detection was contaminated
by the host galaxy (Malesani et al. 2006d).

\item GRB~070223 : the afterglow of that burst was confirmed in the 
infrared band (Castro-Tirado et al. 2007, Rol et al. 2007) and found
to be very faint in the optical band, close to the deep limit of our
observations.

\end{itemize}

In summary, of the 10 afterglows discussed above, 7 were detected in
the infrared bands with very red colours and 3 in the optical band. Of
those latter 3 optical afterglows, two (GRB~060602A and GRB~061006)
had a magnitude below our limiting magnitude in the same band at the
same time. For the remaining one (GRB~060121) our observations
performed $\sim 2.5$ hours after the burst support a probable
high-redshift nature for this event.

\section{Discussion}

\subsection{A Comparison of X-ray and Optical Light Curves}

For a simple visual comparison we show in Figure~\ref{figxopt} the X-ray
light curves (where data were available, from Evans et al. 2007
\footnote[2]{For those bursts where no exact conversion factor from 
count rate to observed flux (0.3-10 keV) was available we assumed the
mean value of $5\times10^{-11}$~erg~cm$^{-2}$~s$^{-1}$.}), together
with the optical light curves, including all published data. When
optical and X-ray data cover the same time interval we superimposed on
the data a simple power-law fit, to better understand the temporal
decay behaviour. As can be seen, the light curves in the two bands do
not follow the same temporal decay for all the GRBs. For the majority
of the bursts the behaviour in the X-ray and optical bands is
different, especially at early times where in the X-ray band the
temporal decay is steep, showing the hints of large flare activity.

\subsubsection{Blastwave Physics from Light Curves Breaks}

In the standard fireball model, observed afterglow emission is
synchrotron radiation from a quasi-spherical relativistic blast wave
(forward shock) that propagates into the homogeneous or wind-like
ambient medium. The model can give clear predictions of the shape of
light curves at different frequencies. Here we test this model
comparing the theoretical expectations with the observed light curves
properties. We take into account simple modifications to the standard
model if needed. The modifications are energy injection into the blast
wave in the afterglow phase ($L \propto t^{-q}$) and a generalized
wind environment ($\rho \propto R^{-s}$).  The injection would modify
the bast wave dynamics as long as $q<1$ and $q=1$ corresponds to the
case without energy injection. The possible theoretical values for the
temporal decay index $\alpha$ and the spectral index $\beta$ as
functions of the electron spectral index $p$ are summarized in Table
3.

Traditionally, optical data from ground based telescopes alone have
been used to establish the presence of achromatic breaks since the
number of pre-{\it Swift} bursts with good simultaneous X-ray and
optical data at any time (early and late) was very small. However,
recent studies of {\it Swift} bursts have shown that many {\it Swift}
GRBs exhibit a well defined steepening of the X-ray light curve while
the optical decay continues to be described by a single un-broken
power law (Panaitescu et al. 2006). In some cases the decay is a
straight power law at all times with no breaks either in the optical
or in the X-ray band (Mundell et al. 2007b), while in some bursts the
break is observed only in the optical but not in the X-ray band.

In this paper we study breaks in X-ray and optical light curves in the
decay phase. Immediately after the prompt emission, some light curves
show a peak or flare features which are likely to be due to central
engine activity or reverse shock emission. We concentrate our
discussion on the study of simpler forward shock emission. The bursts
in our sample (see Figure~\ref{figxopt}) can be divided into four main
classes, depending on the presence or not of a break in the optical
curve (see Figure~\ref{figc2}):

\begin{itemize}

\item Class A : no break in the optical or in the X-ray band;
\item Class B : no break in the optical band, break in the X-ray band;
\item Class C : break in the optical band, no break in the X-ray band;
\item Class D : break in the optical and in the X-ray band.

\end{itemize}

In the cartoon shown in Figure~\ref{figc2} we show the shape of the
four classes. This classification is based on the available data in
the two bands: {\it Swift}-XRT for the X-ray light curve and our
telescopes, GCNs and published data for optical light curves. There is
clearly the possibility of additional breaks in the period not covered
by the observations both in the optical and in the X-ray band. The
possible mechanisms that can produce a break in the observed light
curves in the decay phase can be summarised as follow:

\begin{enumerate}

\item the cooling break $\rightarrow$ chromatic break;

\item cessation of energy injection $\rightarrow$ achromatic break;

\item jet break $\rightarrow$ achromatic break;

\item change in the ambient distribution $\rightarrow$ chromatic or achromatic break;

\item additional emission component (reverse shock, late central engine activities, SN-component, host galaxy contamination) $\rightarrow$ chromatic break;

\end{enumerate}

In the epoch under investigation in this paper ($10^{2} - 10^{6}$ s),
the emission process is in the slow cooling regime. One of the most
natural explanations for a break in a light curve is the cooling break
(i): on the passage of the cooling frequency through the observation
band, the light curve steepens by $\delta \alpha = 1/4$. The
steepening happens in the X-ray band first and in the optical band
later for a homogeneous medium, while it occurs in the optical band
first and in the X-ray band later for a wind-like medium. Note that
the cooling frequency increases in time for the wind-like medium.  The
cessation of energy injection into the blast wave (ii) or a jet break
(iii) causes a change in the hydrodynamics of the blast wave,
producing achromatic breaks in the light curves. In the post-break
phase, the optical and X-ray decay indices could be the same or
different by $\alpha = 1/4$ for mechanism (ii). The decay indices in
the two bands should be the same for a jet break (iii). It should be
noted that for the mechanism (iv) the afterglow emission above the
cooling frequency does not depend on the ambient matter density. The
break is achromatic if the cooling frequency is located above the
X-ray band, while it is chromatic if the cooling frequency lies
between the two bands. The mechanisms (v), especially reverse shock
emission and late time internal shocks, are generally believed to be
relevant only at early times and their contribution to the shape of
the light curves becomes negligible at late times.

Temporal and spectral properties of all the bursts in the sample are
reported in Table 1. Assuming the standard fireball model and a
homogeneous or wind-like circumburst medium it is possible to derive
the closure relations between the temporal decay index ($\alpha$) and
the spectral slope ($\beta$) in order to satisfy the models
(e.g. Zhang et al. 2006). In Figure~\ref{figab} the comparison between
observed properties and model expectations is shown in the optical
(left panel) or in the X-ray (right panel) band. The two bursts
indicated on the left panel of Figure~\ref{figab} (GRB~060108 and
GRB~060210) are the ones that, based on the optical data, deviate the
most from the standard model. In the case of GRB~060108 the spectral
optical analysis (without accounting for extinction) gives a steep
value for $\beta_{\rm O}$, hard to explain for the standard
model. When extinction is included a shallower value is found
($\beta_{\rm OX} \sim 0.5$) more in agreement with the observed
spectral energy distribution (Oates et al. 2006). For GRB~060210, the
spectral analysis reveals a large difference between $\beta_{\rm X}$,
$\beta_{\rm O}$ and $\beta_{\rm OX}$, probably due to a large amount
of extinction which is difficult to evaluate. If again we consider the
value found from the fit of the multiband spectral energy distribution
a value in agreement with the model is found (see Curran et al. 2007
for detailed analysis of this burst). Using temporal and spectral
information (when available), it is now possible to constraint the
value of electron spectral index ($p$) for different bursts and study
each burst belonging to our four classes in the context of the
standard fireball model.

\subsubsection{{\it Class A - no breaks}} 

For 10 bursts (GRB~050502A, GRB~060203, GRB~060204B, GRB~060418,
GRB~060512, GRB~061007, GRB~061110B, GRB~070208, GRB~070411 and
GRB~070419A) no break is observed in the optical or X-ray bands, but
in general, the decay indices of the optical and the X-ray light
curves are different.

The simplest explanation for the difference is that the cooling
frequency is situated between the optical and the X-ray band. If this
is the case, the difference is $\delta \alpha = 1/4$ as we have
discussed in the previous section. If there is energy injected into a
blast wave $L \propto t^{-q}$, the difference is given by $\delta
\alpha = (2-q)/4$ (see Table~3). The X-ray light curve is
steeper by $\delta \alpha$ than the optical in the homogeneous medium,
while the optical one is steeper by $\delta
\alpha$ in the wind-like medium. In the generalized wind-like medium
case $\rho \propto R^{-s}$ (with no energy injection), optical
afterglow decays faster by $\delta \alpha = (3s-4)/(16-4s)$ than the
X-ray afterglow (Monfardini et al. 2006).

\begin{itemize}

\item GRB~050502A : the decay index of the X-ray afterglow is not well 
determined, and late time {\it Swift} observations give only a lower
limit for $\alpha_{\rm X}$. Since the optical decay is shallower,
$\nu_{\rm c}$ should lie between the two bands and the homogeneous
ambient medium is favored. The electron spectral index is given by $p
= 2\beta_{\rm X} = 2.6\pm0.3$. The theoretical values $\alpha =
3(p-1)/4 = 1.2$ and $\beta = (p-1)/2 = 0.8$ are in good agreement with
the observations (Guidorzi et al. 2006).

\item GRB~060203 : the X-ray light curve is steeper by $\delta \alpha 
\sim 0.2$ than the optical light curve. It indicates the uniform ambient 
medium and $\nu_{\rm O}<\nu_{\rm c}<\nu_{\rm X}$ during the power law
decay phase. The theoretical estimates $\alpha_{\rm O} \sim 0.75$,
$\alpha_{\rm X} \sim 1.0$ and $\beta_{\rm X} \sim 1.0$ ($p \sim 2.0$)
can explain well the observations ($\alpha_{{\rm O},1} = 0.74\pm0.13$,
$\alpha_{{\rm X},1} = 0.94 \pm 0.05$ and $\beta_{\rm X} = 0.9 \pm
0.2$).

\item GRB~060204B : the X-ray afterglow decays faster by $\delta \alpha 
= 0.62$ than the optical afterglow. The difference $\delta \alpha$ is
larger than the value $\delta \alpha = 1/4$ for a simple model in
which the optical and X-ray bands are in different spectral domain
(i.e. $\nu_{\rm O}<\nu_{\rm c}<\nu_{\rm X}$). Since X-ray light curve
is steeper than the optical light curve, wind-like medium model does
not work. Although constant energy injection (q=0) gives a close value
$\delta \alpha = 0.5$ we do not expect the observed steep decay
$\alpha_{{\rm O},1} = 0.73$ and $\alpha_{{\rm X},2} = 1.35$ for such
significant energy injection. In the early phase flares are noticeable
in the X-ray light curve. Late internal shock emission might dominate
X-ray band at later times as well. Superposed flares might steepen the
X-ray light curve. Another possible explanation is that fluctuations
in the ambient medium produce bumps in the late optical light curve
(this interpretation was already proposed by Guidorzi et al. 2006 to
explain the bump in the light curve observed for GRB~050502A). If the
cooling frequency lies between the two bands, the bumps are produced
only in the optical light curve. Since the optical observations are
very sparse for this afterglow,a bump in the optical light curve might
take the decay index shallower than the real value.

\item GRB~060418 : the X-ray light curve is steeper by $\delta \alpha = 
0.25$ than the optical light curve in the late decay phase. This
indicates an uniform ambient medium with $\nu_{\rm O}<\nu_{\rm
c}<\nu_{\rm X}$ during that phase. Temporal observed values
($\alpha_{{\rm O},1} = 1.19$, $\alpha_{{\rm X},1} = 1.44$) are in
agreement with the theoretical expectation of a simple model with a
value for the spectral index of $p \sim 2.6$ ($\alpha_{\rm O} = 1.20$,
$\alpha_{\rm X} = 1.45$).

\item GRB~060512 : the X-ray emission decays faster by $\delta \alpha = 
0.38$ than the optical emission. The difference $\delta \alpha$ is not
consistent with the simplest scenario ($\delta \alpha$ = 1/4). The
energy injection (q=0.48) in homogeneous ambient can account for the
large value of $\delta \alpha$. However, even with $p=2\beta_{\rm
X}=2.2$, the maximum value allowed from the observed spectral index
$\beta_{\rm X}=0.93\pm0.18$, the expected decay indices $\alpha_{\rm
O} = [(2p-6)+(p+3)q]/4 = 0.22$ and $\alpha_{\rm X} = [(2p-4)+(p+2)q]/4
= 0.60$ (Zhang et al. 2006) are much shallower than the observed
values. X-ray flares might make the X-ray decay index larger than the
real decay index of the blast wave emission.

\item GRB~061007 : the afterglow of this burst is very bright in optical 
and X-ray, the decay is a straight power law from early time (there is
the hint of a rise in the optical not mirrored in the X-ray) since
late times. A comprehensive multiwavelength analysis of this burst is
presented in Mundell et al. (2007b): the evolution of the afterglow
can be explained in the context of the fireball model with $\nu_{\rm
m} < \nu_{O} < \nu_{\rm X} < \nu_{c}$ for the entire $10^6$~s period
covered by the observations.

\item GRB~061110B : for that burst the optical light curve decays faster 
by $\delta \alpha = 0.20$ than the X-ray, indicating that during the
observations $\nu_{\rm c}$ is located between the two bands in a
wind-like medium. The X-ray data imply a value of $p \sim 2.5$ and the
expected value for the optical temporal decay $\alpha_{\rm O} \sim
1.63$ is in good agreement with that observed ($\alpha_{{\rm O},1} =
1.64\pm0.08$).

\item GRB~070208 : the X-ray afterglow decays faster by $\delta \alpha 
= 0.87$ than the optical afterglow. Again this difference is much
larger than the value $\delta \alpha = 1/4$ for a simple model,
similar to the case of GRB~060204b. Also the energy injection model
can not account for the larger value $\delta \alpha = 0.87$. Beyond
the standard model, possible explanations of such a large difference
are X-ray flares (late time internal shocks) which make the X-ray
steeper coupled with energy injection which makes the optical decay
shallower.

\item GRB~070411 : the X-ray afterglow decays faster by $\delta \alpha 
= 0.20$ than the optical afterglow. It indicates the uniform ambient
medium and $\nu_{\rm O}<\nu_{\rm c}<\nu_{\rm X}$ during the power law
decay phase. If this is the case, the value of p derived from the
X-ray data ($p \sim 2.2$) imply a value of the decay indices
$\alpha_{\rm O} \sim 0.9$ and $\alpha_{\rm O} \sim 1.2$, well in
agreement with the observed values ($\alpha_{{\rm O},1} = 0.92\pm0.04$
and $\alpha_{{\rm X},1} = 1.12\pm0.03$).

\item GRB~070419A : for this burst the temporal decay in the two bands 
is again very similar, but this is true only at late times ($\delta
\alpha_{A} = 0.06$), while at early times the shape in the two bands
is very different: a very steep decay in the X-ray ($\alpha \sim 2.8$)
and a possible broad re-brightening in the optical. Even in the late
power law phase, no closure relations for the simple models can
reconcile the observed value $\alpha_{\rm X} - 3\beta_{\rm X}/2 =
-1.6$. Assuming $\nu_{\rm X},\nu_{\rm O}>\nu_{\rm c}$ (then the
emission does not depend on the ambient medium) we obtain $p=2.9$ from
$\beta_{\rm X}$. The observed decay indices in the two bands
($\alpha_{\rm O} \sim \alpha_{\rm X} \sim 0.6$) and the closure
relation could be explained if there is significant energy injection
($q=0.12$). The total injected energy increases by a factor $(3
\times 10^{6}/3500)^{(1-q)} = 380$ between the break time $t_{{\rm
X},break} \sim 0.04$~days$\sim 3500$~s and the end of the observations
$\sim 3 \times 10^{6}$~s. This could contradict with the energy budget
of the central engine (solar mass scale).

\end{itemize}

\subsubsection{{\it Class B - break in the X-ray band only}} 

A steepening in the X-ray decay slope is observed in eight bursts of
our sample (GRB~050713A, GRB~060108, GRB~060210, GRB~060510B,
GRB~060927, GRB~061121, GRB~070420 and GRB~070714B), while the rate of
the optical decay remains constant.

A simple explanation of this behaviour could be the passage of the
cooling frequency through the X-ray band. For a homogeneous ambient
medium, the decay indices of the optical and X-ray light curves should
be the same in the pre-break phase, with only the X-ray light curve
steepening due to the passage of $\nu_{\rm c}$. In contrast, for
wind-like medium the optical light curve is steeper than the X-ray
light curve in the pre-break phase and the decay indices in the two
bands become the same after a break in the X-ray light curve. For the
eight bursts in that class, the observed steepening $\delta \alpha$ is
always larger than the value $\delta \alpha = 1/4$ expected in the
simplest scenario.

\begin{itemize}

\item GRB~050713A : the X-ray emission decays faster by $\delta \alpha
\sim 0.7$ than the optical in the post-break phase. The decay indices
in the pre-break phase are also significantly different from each
other ($\delta \alpha = 0.54$). We cannot explain the break (and the
behaviour of the light curves in the two bands) by the cooling break
even if the energy injection and generalized wind-like medium are
assumed. The cessation of the energy injection and a jet break also
cannot account for the observed break because of their achromatic
nature. The most likely explanation is that X-ray flares due to late
internal shocks shapes the X-ray light curve (there are notable
fluctuations in the X-ray light curve). Although the optical light
curve is poorly sampled, the rather shallow observed decay
($\alpha_{{\rm O},1} = 0.63\pm0.04$) might indicate energy injection
into the blast wave.

\item GRB~060108 : the decay indices of the optical and X-ray light 
curves are almost the same in the pre-break phase, but the steepening
in the X-ray band ($\delta \alpha = 0.69$) is too large to be
explained in a simple cooling break model (also reported by Oates et
al. 2006). Since there are no optical observations at late times,
achromatic break mechanisms are also applicable to this event. The
X-ray decay after the break ($\alpha_{{\rm X},2} = 1.15$) is too
shallow for the jet break model in which the electron spectral index
should be equal to the post-break decay index ($p=\alpha_{{\rm X},2}$
in that case). Considering the shallow pre-break decay in the X-ray
and optical, the probable explanation is the energy injection ceasing.

\item GRB~060210 : the optical light curve is complex. The early flat 
portion might be due to energy injection or to the passage of the
typical frequency through the optical band, and there is a hint of a
late break. Since we are interested in breaks in the decay phase, we
discuss the intermediate power law part of the optical light curve
together with X-ray observations. Before the X-ray break the optical
afterglow decays faster ($\delta \alpha = 0.15$) than the X-ray
afterglow. This could indicate that the cooling frequency lies between
the two bands and that the ambient medium is wind-like. Since after
the break the X-ray light curve becomes steeper than the optical, the
steepening is not explained by a cooling break even if we consider
energy injection or a generalized wind-like medium. The post-break
X-ray index ($\alpha_{{\rm X},2} = 1.31$) is too shallow to be
explained in the jet break model. In the model of cessation of energy
injection the difference of the decay indices in the two bands
($\nu_{\rm O}<\nu_{\rm c}<\nu_{\rm X}$) is $\delta \alpha = (2-q)/4$in
the pre-break phase (during the energy injection) and the steepening
in the X-ray light curve (ceasing of energy injection) is given by
$\delta \alpha_{\it break} = (p+2)(1-q)/4$. The observed difference
$\delta \alpha = 0.15$ is smaller than the expected value in the
energy injection model ($q$ should be smaller than unity). The X-ray
spectrum $\beta_{\rm X}=1.14$ and the observed steepening $\delta
\alpha_{\it break} = 0.43$ require $p=2.8$ and significant energy
injection ($q=0.60$), for which the decay indices are expected to be
$\alpha_{{\rm O},1} = 1,13$, $\alpha_{{\rm X},1} = 0.78$,
$\alpha_{{\rm X},2} = 1,21$. This could marginally explain the
observations. A more plausible explanation is that X-ray flares due to
late central engine activity shape the X-ray light curve.

\item GRB~060510B : in the pre-break phase, the decay indices in 
the two bands are the same if the prompt emission and X-ray flares at
early times are ignored. Considering the shallow pre-break decay in
the X-ray and optical bands, a possible explanation for the X-ray
break is the cessation of energy injection. If this is the case, the
optical light curve should have a break at the same time, although
there are no optical observations at late times.

\item GRB~060927 :  the optical light curve in the R-band does not 
show a simple decay; $\alpha_{{\rm O},1} > \alpha_{{\rm X},1}$ and
$\delta \alpha > 1.0$. There is a possible flat phase in the optical
between $10^2$ and $10^3$ s. A possible explanation could be late
energy injection but this behaviour is not mirrored in the X-ray
band. The observed characteristics of this burst are difficult to
explain in the context of the standard model.

\item GRB~061121 : Page et al. (2007) found that the decay indices 
in the optical and X-ray bands are consistent, within the
uncertainties, with the (possible) presence of an achromatic
break. However, the optical data after the break, as reported in our
Figure~\ref{figxopt3}, are in very good agreement with the same power
law decay index observed before the break ($\alpha_{{\rm O},1} =
0.83$), without requiring any break in the optical. The chromatic
nature of the break excludes the possibilities of a jet break and
cessation of energy injection for the explanation of the break in the
X-ray light curve. Since the X-ray decay index evolves from almost
zero (flat) to a steep value ($\alpha_{{\rm X},2} = 1.58$), which is
much larger than the optical decay index ($\alpha_{{\rm O},1} =
0.83$), neither cooling break models nor change of the ambient density
distribution (e.g. change from ISM medium to wind-like medium during
the propagation of the blast wave) can explain the evolution of the
X-ray light curve. The emission from late central engine activities
could mask the X-ray radiation from the forward shock. The darkness
$\beta_{\rm OX} = 0.53$ might be due to the bright additional X-ray
emission from the central engine.

\item GRB~070420 : the observed data are not consistent with any closure
relation for the standard model. The behaviour of the X-ray and the
optical light curves looks quite different. However, the sparse
observations can not exclude the presence of a flat phase in the
optical light curve as visible in the X-ray. If this is the case, the
ceasing of energy injection could explain the X-ray and optical
observations.

\item GRB~070714B : similarly to GRB~070420, the behaviour or the
X-ray and optical light curves cannot be explained in the standard
model if a single power law is assumed for the decay of the optical
light curve. Although the optical data are too sparse to firmly
constrain the behaviour in that band, but a possible initial flat
phase may be present (like in the X-ray band) followed be a standard
power law decay. Like the previous case, the ceasing of energy
injection could explain the X-ray and optical behaviours.

\end{itemize}

\subsubsection{{\it Class C - break in the optical band only}} 

In five cases a change of decay index is detected only in the optical
light curve and not in the X-ray band (GRB~041006, GRB~041218,
GRB~051111, GRB~060206 and GRB~061126).

A possible explanation for these breaks is the passage of $\nu_{\rm
c}$ through the optical band. If a homogeneous medium is assumed, the
decay indices of the optical and the X-ray light curves should be the
same in the post-break phase, with only the optical light curve
steepening with the passage of $\nu_{\rm c}$. If a wind-like ambient
medium is assumed, the decay indices in the two bands are the same in
the pre-break phase and the optical light curve is steeper than the
X-ray after the break.

\begin{itemize}

\item GRB~041006 : counting the rather large error in the value of 
$\alpha_{\rm X}$, the X-ray decay and the post-break optical decay
indices could be considered as almost the same. If the cooling break
model is assumed to explain the steepening of the optical light curve,
a homogeneous ambient medium is favored, because the decay indices in
the two bands are the same in the post-break phase. The drastic
steepening in the optical light curve ($\delta \alpha = 0.53$)
requires almost constant energy injection ($q\sim0$). It is hard to
achieve the steep decay in the post-break phase ($\alpha_{{\rm O},2} =
1.12$) with such a massive energy injection. The cooling break model
does not work. Since there are no X-ray observations at early times,
achromatic break models are acceptable. The post-break optical index
is too shallow to consider a jet break.

Next we consider the cessation of energy injection. From the observed
spectral indices ($\beta_{\rm O} \sim \beta_{\rm X} \sim 1.0$), the
two bands should be in the same spectral domain and we obtain $p\sim3$
for $\nu_{\rm m}<\nu_{\rm obs}<\nu_{\rm c}$ or $p\sim2$ for $\nu_{\rm
c}<\nu_{\rm obs}$. The closure relations are marginally satisfied in
both cases. The ceasing of energy injection causes steepening in a
light curve. If the observation band is above $\nu_{\rm c}$ the flux
does not depend on the ambient medium and the steepening is $\delta
\alpha = (p+2)(1-q)/4$. If $\nu_{\rm m}<\nu_{\rm obs}<\nu_{\rm X}$ the
steepening is $\delta \alpha = (p+3)(1-q)/4$ for a homogeneous ambient
medium and $\delta \alpha = (p+1)(1-q)/4$ for a wind-like ambient
medium. For the combination of $\nu_{\rm m}<\nu_{\rm obs}<\nu_{\rm c}$
and homogeneous ambient medium, the initial energy injection is
mildest $q=0.47$ and the expected values $\alpha_{\rm pre-break} \sim
0.71$, $\alpha_{\rm post-break} \sim 1.5$ and $\beta \sim 1.0$ could
be consistent with the observations.

\item GRB~041218 : there is only a late time observation for the X-ray 
band, neither $\alpha_{\rm X}$ nor $\beta_{\rm X}$ are constrained
from the observations. The optical spectral index $\beta_{\rm O}$ is
also not available. The break in the optical light curve ($\delta
\alpha =0.22$) could be explained in many models including the cooling
break.

\item GRB~051111 : since the X-ray decay index $\alpha_{{\rm X},2} =
1.60$ is incoincident with both the pre-break ($\alpha_{{\rm O},1} =
0.82$) and the post-break ($\alpha_{{\rm O},2} = 1.0$) optical decay
index, cooling break models cannot account for the optical break even
if energy injection is considered. The fact that the X-ray emission
decays faster than the optical emission rules out a wind-like ambient
medium scenario (and scenarios related to the wind medium). No X-ray
observations are available before the optical break and the break
might be achromatic. The jet break is unlikely because the afterglow
decays with significantly different rates in the two bands after the
break. The difference of the spectral indices in the two bands
indicates that the two bands are in different spectral domains
(i.e. $\nu_{\rm m}<\nu_{\rm O}<\nu_{\rm c}<\nu_{\rm X}$). This
highlighted by the spectral energy distribution analysis presented in
Guidorzi et al. 2007. Using the observed $\beta_{\rm O} = 0.76$ (for
which the error is smaller than in $\beta_{\rm X}$), we obtain
$p=2\beta_{\rm O}+1=2.5$. If the energy injection rate changes from
$q_{1}$ to $q_{2}$ at the break, the steepening of the optical decay
is given by $\delta \alpha_{\rm O} = (p+3)(q_{2}-q_{1})/4 = 0.18$ and
the difference of the decay indices is $\delta \alpha = (2-q_{2})/4 =
0.6$ in the post-break phase. The resulting $q_{1}$ and $q_{2}$ are
negative and unphysical, with which the predicted temporal and
spectral indices deviate largely from the observed values. The energy
injection model also does not work.

\item GRB~060206 : the X-ray light curve is consistent with a single 
unbroken power law, achromatic break models (i.e. jet break and
ceasing of energy injection) are ruled out. Since the X-ray decay
index ($\alpha_{{\rm X},1} =1.30$) is different from both the
pre-preak ($\alpha_{{\rm O},1} =0.93$) and the post-break
($\alpha_{{\rm O},2} =1.83$) optical decay indices, the cooling break
models can not account for the optical break even if the energy
injection or generic wind-like ambient medium is considered. The
observed behaviour of the light curves might be due to a transition in
the ambient matter distribution (Monfardini et al. 2006): a blast wave
initially propagates into a constant medium and then it breaks out
into a wind-like medium. Note that X-ray emission does not depend on
the ambient matter density (and its distribution) as long as the X-ray
band is above the cooling frequency, and that the optical emission
(below $\nu_{\rm c}$) reflects the change in the ambient matter
distribution.

\item GRB~061126 : for this burst no break is visible in the X-ray band,
while in the optical the transition is from steeper to shallower decay
index. This behaviour at early times can be explained as the
contribution from the reverse shock component. The detailed study by
Gomboc et al. (2008) shows its inconsistency with the standard
fireball model (the steeper decay in the X-ray band and the large
ratio of X-ray to optical flux).

\end{itemize}

\subsubsection{{\it Class D - break in both bands}} 

The bursts belonging to this class are those showing a break in both
the optical and the X-ray light curves in their decay phases. In
general, breaks in the two bands occur at different times, both
chromatic and achromatic breaks are considered. Surprisingly, for only
one burst do we observe a break in both bands, although GRB~060210
might be classified in this case if we take seriously the last optical
data points.

\begin{itemize}

\item GRB~050730 : if a jet break is responsible for the steep X-ray
decay, the electron energy index is $p=\alpha_{{\rm X},2}=2.37$.
Given that X-ray band is below the cooling frequency, the observed
X-ray spectral index $\beta_X= 0.73\pm0.07$ is consistent with the
model prediction $\beta=(p-1)/2=0.69$.  However, the much shallower
optical decay ($\alpha_{{\rm O},2}=1.55\pm0.08$) is inconsistent with
the jet break model. As pointed out by Pandey et al. (2006), the
possibility of a contribution from the host galaxy or an associated SN
to the late time optical afterglow can be ruled out considering the
high redshift of the burst ($z=3.967$). Since after a jet break a
forward shock emits photons practically at a constant radius (the
exponential slowing down; Sari, Piran $\&$ Halpern 1999), fluctuations
in the ambient medium do not seem to affect the decay rate of the
emission. Even if there is an effect, both light curves should become
shallower (or steeper) in the same way because both optical and X-ray
bands are in the same spectral domain ($\nu_O,\nu_X < \nu_c$). Energy
injection into a forward shock also cannot explain the shallow optical
decay because of the same reason. Additional emission components,
e.g. the two component jet model or late time internal shocks might
make the optical decay slower. The early shallow decay phase observed
in the optical and X-ray light curves could be explained by energy
injection (Pandey et al. 2006), though the ceasing of the injection
should happen around the time of the jet break.

\end{itemize}

\subsubsection{Summary of Light Curves Breaks}
 
From this analysis of the optical and X-ray light curves we conclude
that in our sample of 24 optical GRB afterglows: 15 bursts can be
explained in the context of the standard fireball model (with
modifications: energy injection or variation in the ambient matter
distribution); while for the remaining 9 bursts:

\begin{itemize}

\item Class A : GRB~060204B, GRB~060512, GRB~070208, GRB~070419A
\item Class B : GRB~050713A, GRB~060927, GRB~061121
\item Class C : GRB~051111, GRB~061126

\end{itemize}

the observed data are inconsistent with the predictions of the
standard model.

\subsection{Rest Frame Properties}

In Figure~\ref{figall2} we translate the observed magnitudes of the
optical afterglows into the rest-frame luminosity (left panel). The
subscript $t$ used in this section refers to the time in the rest
frame of the GRBs. We assumed a standard cosmology (defined in Section
1) and we include any correction expected from the distance of the
event ($z$) and its spectral properties ($\beta$) in order to report
all the observed quantities in the rest frame of the GRB. We corrected
the optical magnitude for Galactic extinction using the reddening maps
of Schlegel et al. (1998) and we applied the {\it k}-correction to
take into account the fact that sources are observed at different
redshifts (${\it k} = -2.5~log (1+z)^{(\beta-1)}$). We do not correct
for the host galaxy dust absorption. From this analysis we excluded :
1) the bursts for which no spectroscopic redshift was available
(GRB~041218, GRB~050713A, GRB~060108, GRB~060203 and GRB~060204B); 2)
GRB~060510B, for which there is a value of the spectroscopic redshift
but the optical light curve is sparsely sampled. After this selection
our sub-sample totals 16 objects.

Even in the rest frame of the burst, starting the observations about
0.5 minutes after the burst event, a difference of about 4 orders of
magnitude in luminosity is evident, particularly at early time. This
spread in intrinsic luminosity remains after including all the
available early and late time public data, although there is a hint of
a convergence at later times. It should be noted that our analysis
does not take into account any beaming effect. This collimation
correction is known to reduce the observed spread in luminosity for
bursts (Frail et al. 2001, Panaitescu $\&$ Kumar 2001) but requires a
correct determination of the jet opening angle for each burst based on
an unambiguous identification of a jet break. Identification of such
break times is clearly important but, as discussed in Section 4.1.1 is
non-trivial due to complex light curve properties and requires
well-sampled optical and X-ray light curves from the earliest to the
latest possible times.

In a previous study of optical afterglow light curves, Liang $\&$
Zhang (2006, hereafter LZ06) suggested that the optical luminosity at
t=1 day (source frame time) after the burst shows a bimodal
distribution, with a separation at $L^{*}_{\rm t=1day} = 1.4 \times
10^{45}$~erg~s$^{-1}$. The majority of the bursts in their sample (44
bursts in total) fall into the luminous group (34 bursts with
$L_{peak} > L^{*}_{t=1day}$). Kann et al. 2006 and Nardini et al. 2006
(hereafter K06 and N06 respectively) found a similar result. LZ06
selected t=1 day as the reference time because at this time the light
curves of their sample were better sampled. Moreover they selected
this late time because they were concerned about the possible
contribution of the reverse shock component or additional energy
injection at early times. The result of N06 was obtained in the same
way but extrapolating the luminosity at t=0.5 days (source frame
time).

As our observations have good coverage starting at earlier time
(between 1 and 20 minutes in the GRB rest-frame) we have estimated the
intrinsic optical luminosity at three different times: 10 minutes, 0.5
days and 1 day (source frame time). In the cases presented here we
have confirmed that the reverse shock component does not affect our
analysis. As discussed in the previous section, in only one case
(GRB~061126) do we detect the possible contribution of the reverse
shock at early times. Our early observations at t=10 minutes are more
directly related to the explosion energy during the prompt emission
phase.

In Table 4 we report the mean values for the rest-frame luminosity
calculated at different times. The two classes defined by LZ06 (dim
and lum) are not consistent with a single population, as clear from
their Figure~2.  Our data (at any times) are consistent within the
uncertainties with a single population rather than with two separate
classes. In Fig~\ref{figlum12h} we show the observed luminosity
distribution of our sample extrapolated at 12 hours.The distribution
is well fitted with a single log-normal function with an average of
$29.54\pm0.07$ and a $\sigma$ of $0.67\pm0.05$.

Figure~\ref{figall2} (right panel) shows the luminosity-redshift
distribution for the bursts in our sample. In this figure we over plot
the separation line between the two classes and the highest values for
the redshift of the members of the two groups in the LZ06 sample. As
pointed out by LZ06, a possible reason for the lack of high redshift
members in their dim group is a lack of deep and rapid followup
observations.  The burst population detected by {\it Swift} has a
larger mean redshift and fainter brightness distribution that previous
missions ($<z>_{{\it Swift}} \sim 2.7$, $<z>_{pre-{\it Swift}} \sim
1.5$, Le $\&$ Dermer 2007). This could explain the results of LZ06,
whose sample was based on bursts detected up to August 2005, thus
containing only 7 {\em Swift} bursts.  The study presented by K06 of a
sample of 16 pre-{\it Swift} bursts similarly probed the bright end of
the GRB luminosity function and found similar conclusions to LZ06. K06
found that on average low-redshift afterglows are less luminous than
high-redshift ones, suggesting a bimodal luminosity distribution.
Strong selection effects due to observational bias against
intrinsically faint afterglows at higher redshifts is a likely
explanation for this result. This observational bias is greatly
reduced in our sample thanks to the rapid response and use of red
filters on our ground based telescopes to {\it Swift} triggers (14/16
objects used for this analysis were detected by {\it Swift}).  Our
results show that faint {\em Swift} GRBs at higher redshift are
readily detected with such rapid, deep optical observations in red
filters (Figure~\ref{figall2}, right panel), a region of parameter
space not accessible in the samples of LZ06 or K06.

The population that is not prevalent in the right panel of
Figure~\ref{figall2} is bright bursts at low redshifts, probably due
to the GRB luminosity function such that very luminous bursts rare and
a large survey volume is therefore required to detect them. Other
authors discuss the possibility of two separate luminosity functions
for luminous and underluminous GRBs (Nardini et al. 2007, Kann et
al. 2007, Liang et al. 2007, Chapman et al. 2007) but given the many
complex instrumental selection effects inherent in GRB discovery and
followup, significantly larger samples are required to draw robust
conclusions. Lower {\em Swift} trigger thresholds may provide the
basis for such samples in the future.

\subsection{Dark Bursts}

We consider as an optically dark event, or ``dark burst'' to be a one
that satisfies the definition of Jakobsson et al. (2004), i.e. that
the slope of the spectral energy distribution between the optical and
the X-ray band or spectral index $\beta_{\rm OX}$, is $<$ 0.5. Even
optically detected bursts may be classified as ``dark'', providing
that the optical flux is much fainter than expected from scaling the
X-ray flux (e.g. GRB060108). For all the bursts reported in Table 2 no
optical counterpart was detected by our telescopes. Apart from a few
cases in which our observations were performed at late times, the
majority of GRBs in our sample were observed by our telescopes
reacting rapidly and performing deep, early-time optical observations
($R \sim 21$ at 5 minutes after the trigger, for co-added images). A
late response or poor sensitivity are therefore ruled out as
explanations for non-detections in most cases. What, therefore, is the
explanation for the lack (or faintness) of the optical afterglow for
these 'dark' bursts?  To understand this we analyse the X-ray light
curve of the bursts in our sub-sample observed by the {\it
Swift}-XRT. Using the decay inferred from the fit of the light curves
the X-ray and optical fluxes are extrapolated to a common time. We
assume that the optical light curves for those undetected bursts
follow a power law decay with a slope equal to the mean temporal decay
of the detected afterglows analysed in Section 3 ($<\alpha_{\rm
O}>~\sim 1.1$). Alternatively, in the X-ray band we use the value
derived from the fit of the light curve and reported on Table 2
($\alpha_{\rm X}^{\rm (fit)}$). Three characteristic times t$_{0}$,
t$_{1}$ and t$_{2}$ are used here. The time t$_{2}$ = 11 hours is
chosen for consistency with the dark burst classification of Jakobsson
et al. (2004), while fluxes extrapolated to t$_{1}$ = 1 hour and
t$_{0}$ = 10 minutes exploit our early-time data without compromising
observing sensitivity.

As can be seen in Figure~\ref{figuljako}, the majority of the bursts
are located close to the dark bursts region ($\beta_{\rm OX}<0.5$)
independent of the selected time, ruling out late observation time as
an explanation for the apparent darkness of most bursts. For almost
all the bursts, the evolution of the optical and X-ray flux follows a
line almost parallel to the lines of constant $\beta_{\rm OX}$ (bottom
right panel). This behaviour can be seen, in the optical band, as a
consequence of the assumption of an average decay when making the
extrapolation. Any change in the temporal decay would clearly modify
this behaviour. However, it seems that bursts that are classified as
``normal'' after 10 minutes (bottom left panel) remain in the same
class also after 11 hours (top left panel). At the same time those
bursts that are optically dark soon after 10 minutes belong to the
class of the so-called ``dark bursts'' also after 11 hours.

In only five cases, does the classification of the burst depend on the
time for the extrapolation of the flux. For these five bursts we
report also the errors (including the uncertainties on the flux and on
the $\alpha$ used for the extrapolation) in the bottom right panel of
Figure~\ref{figuljako}. Three bursts are classified as normal bursts
if we extrapolate at t=t$_{2}$ but belong to the class of ``dark
bursts'' if we extrapolate at t=t$_{0}$ (GRB~050124, GRB~060901 and
GRB~070721B, red circles on Figure~\ref{figuljako}); in two cases it
is the contrary (GRB~050504 and GRB~070219, blue circles). In the
latter cases it is clear that the optical flux significantly
suppressed compared to the X-ray flux. This may be due to spectral
evolution of those bursts, but within the uncertainties we cannot rule
out the possibility that nothing changes also for these five bursts.

For 10/39 bursts, as reported in Table 2, an optical/infrared
counterpart was identified by larger optical or IR telescopes, with
counterparts detected primarily at NIR wavelengths. Our observations
rule out the large population of bright optical counterparts that were
predicted, pre-{\it Swift}, to exist and be observable with suitably
rapid followup observations.  Explanations for dark bursts in the era
of rapid followup remain: extinction caused by dust (Galactic or
host), high redshift origin, or both.  In some cases ($\sim 10\%$) the
Galactic absorption along the light of sight in the observing band
(A$_{\rm R}$) for our bursts is significant and may explain the
undetected optical counterpart; however, it is interesting to note
that no IR detections have been reported for these GRBS. The effect of
Ly-$\alpha$ absorption due to an high-redshift event ($z>7$) is
difficult to evaluate but again this effect could be responsible at
least for a fraction of our non-detections (Roming et al. 2006). The
possibility of a rapid temporal decay seems to be the most unlikely:
in fact, assuming that the undetected bursts of Table 2 have a
temporal behaviour similar to the detected afterglows of Table 1 than
the temporal optical decay at early times appears to be shallow, not
steep. Another possibility could be an excess of X-ray emission at
late time; if late-time central engine activity is responsible for the
production of the early X-ray afterglow in some cases, the additional
emission will mask the forward shock X-ray emission and the total flux
in the X-ray band would be higher than the value expected for the
forward shock emission alone. This might explain some dark bursts and
their distribution on the Log~$F_{\rm O}$-Log~$F_{\rm X}$ diagram. A
combination of these mechanisms and others (i.e. intrinsic optical
faintness, low density circumburst medium) may combine to explain the
high number of bursts that remain undetected at optical wavelengths
($\sim 46\%$ in our sample).

\section{Conclusions}

\begin{itemize}

\item We have classified our afterglows sample into four groups based
on breaks in the optical and the X-ray afterglow light curves during
the decay phase. We have used the temporal and spectral properties of
the X-ray and optical afterglows to investigate the blastwave physics
around the break times within the framework of the standard fireball
model (the synchrotron shock model). The majority of the bursts in our
sample (15 out of 24) are consistent with the standard model. However,
for a significant fraction of our sample (9 bursts: GRB~050713A,
GRB~051111, GRB~060204B, GRB~060512, GRB~060927, GRB~061121,
GRB~061126, GRB~070208 and GRB~070419A), the data cannot be explained
by the standard model, even if modifications to the simple model are
made (i.e. energy injection or variation in the ambient matter). A
possible explanation beyond the standard model is that the early X-ray
afterglow is not due to forward shock emission but is instead produced
by late-time central engine activity (e.g. Ghisellini et al. 2007).

\item We have derived the light curves of the optical afterglows in
the source rest frame for those bursts with spectroscopically
confirmed redshifts (i.e. not merely assuming a fixed redshift z=1 for
all bursts). The optical luminosity function measured at $t=10$~mins
and the corresponding distributions for light curves extrapolated to
$t=12$~hours and 1 day are uni-modal, showing no evidence for the
bi-modality suggested by previous authors.  A fit of the distribution
at 10 minutes with a single log-normal yields an average and a sigma
values of $\log L$(erg s$^{-1}$)$=46.55\pm0.18$ and
$\sigma=1.23\pm0.15$, respectively. Liang $\&$ Zhang (2006) reported a
bimodal distribution of optical luminosity at t= 1 day. Two recent
studies on the afterglows of {\it Swift}-era GRBs (Kann et al. 2007
and Nardini et al. 2007) also suggested a clustering of optical
afterglow luminosities at one day and 12 hours (already found by Kann
et al. 2006, and Nardini et al. 2006), showing again a bi-modality in
the luminosity distribution. This discrepancy may be explained by our
ability to detect fainter GRBs at high redshift; in future, larger
samples covering a wide range of GRB luminosities (possibly
facilitated by lower triggering thresholds on {\it Swift}) will
provide stronger tests for the existence of separate classes of GRBs.

\item By comparing X-ray flux densities and optical upper limits, we
have shown that the majority of non-detections in our sample should be
classified as dark bursts. The rapid response of our telescopes to
real-time localisations from {\it Swift} show that there remains a
significant number of genuinely dark GRB afterglows and rapid optical
temporal decay at early time is ruled out as an explanation for
failure to detect optical afterglows at later time. Of our 39
non-detections, ten afterglows were identified by other facilities,
primarily at NIR wavelengths, demostrating a small population of
bursts in high density host environments. The lack of optical/IR
afterglows for the remaining 29 bursts may be due to effects such as
high levels of extinction (Galactic or host), circumburst absorption,
Ly-$\alpha$ absorption due to high-redshift or low-density
environments suppressing production of optical sychrotron (or a
combination of effects). Alternatively, we suggest that if late-time
central engine activity is responsible for the production of the early
X-ray afterglow emission in some cases, the additional emission will
mask the simultaneous, but fainter forward shock X-ray emission and
result in an observed X-ray flux that is larger than expected from
forward shock emission alone. This might explain some dark bursts.
 
\end{itemize}

\acknowledgements

AM acknowledge founding from the Particle Physics and Astronomy
Research Council (PPARC). CGM acknowledges financial support from the
Royal Society and Research Councils UK. The Liverpool Telescope is
operated by Liverpool John Moores University at the Observatorio del
Roque de los Muchachos of the Instituto de Astrofisica de
Canarias. The Faulkes Telescopes, now owned by Las Cumbres
Observatory, are operated with support from the Dill Faulkes
Educational Trust. This research has made use of the NASA/IPAC
Extragalactic Database (NED) which is operated by the Jet Propulsion
Laboratory, California Institute of Technology, under contract with
the National Aeronautics and Space Administration.


\clearpage

\clearpage

\clearpage

\begin{figure} \centering 
\includegraphics[width=15cm,height=9.5cm]{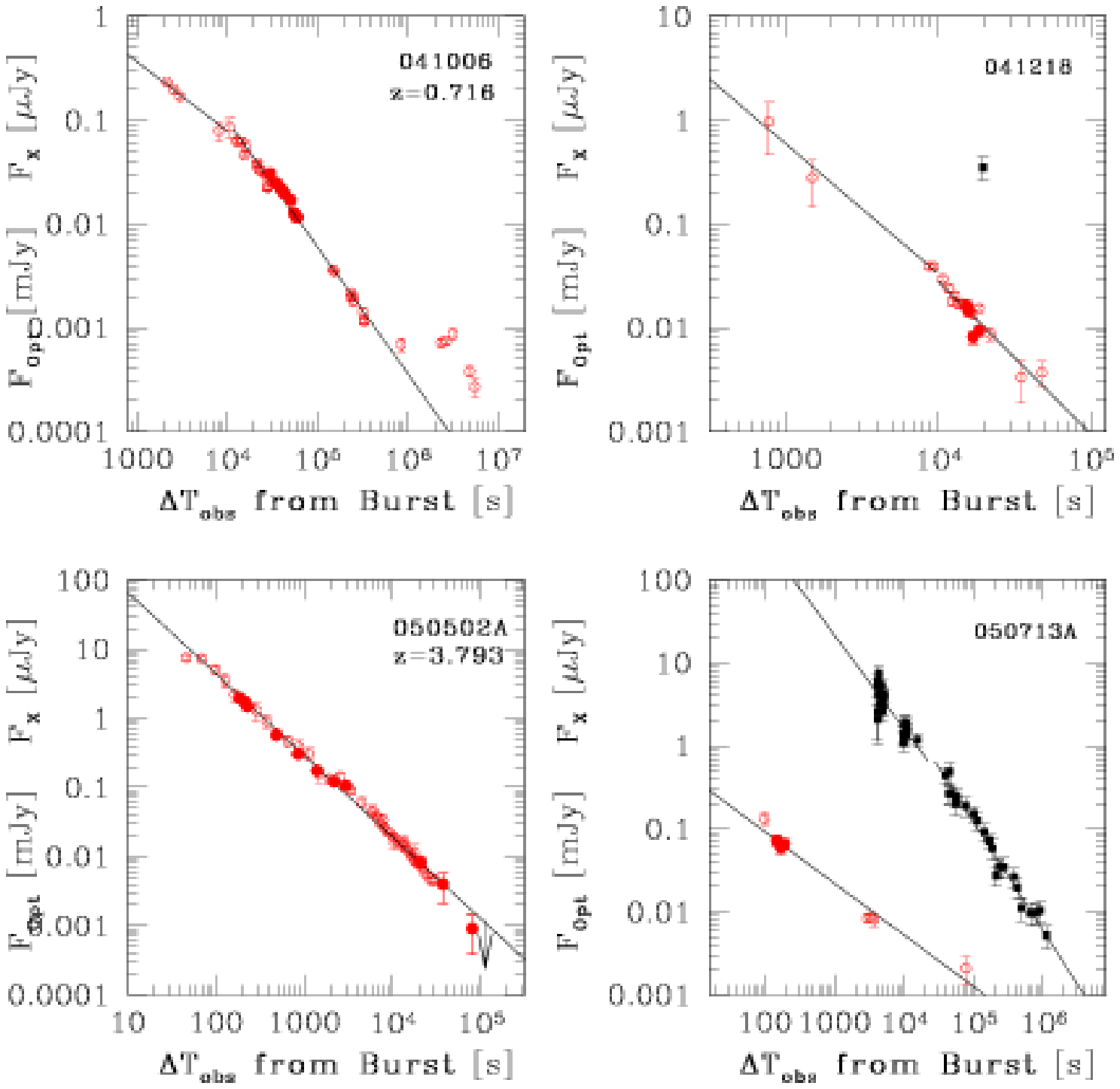}
\includegraphics[width=15cm,height=9.5cm]{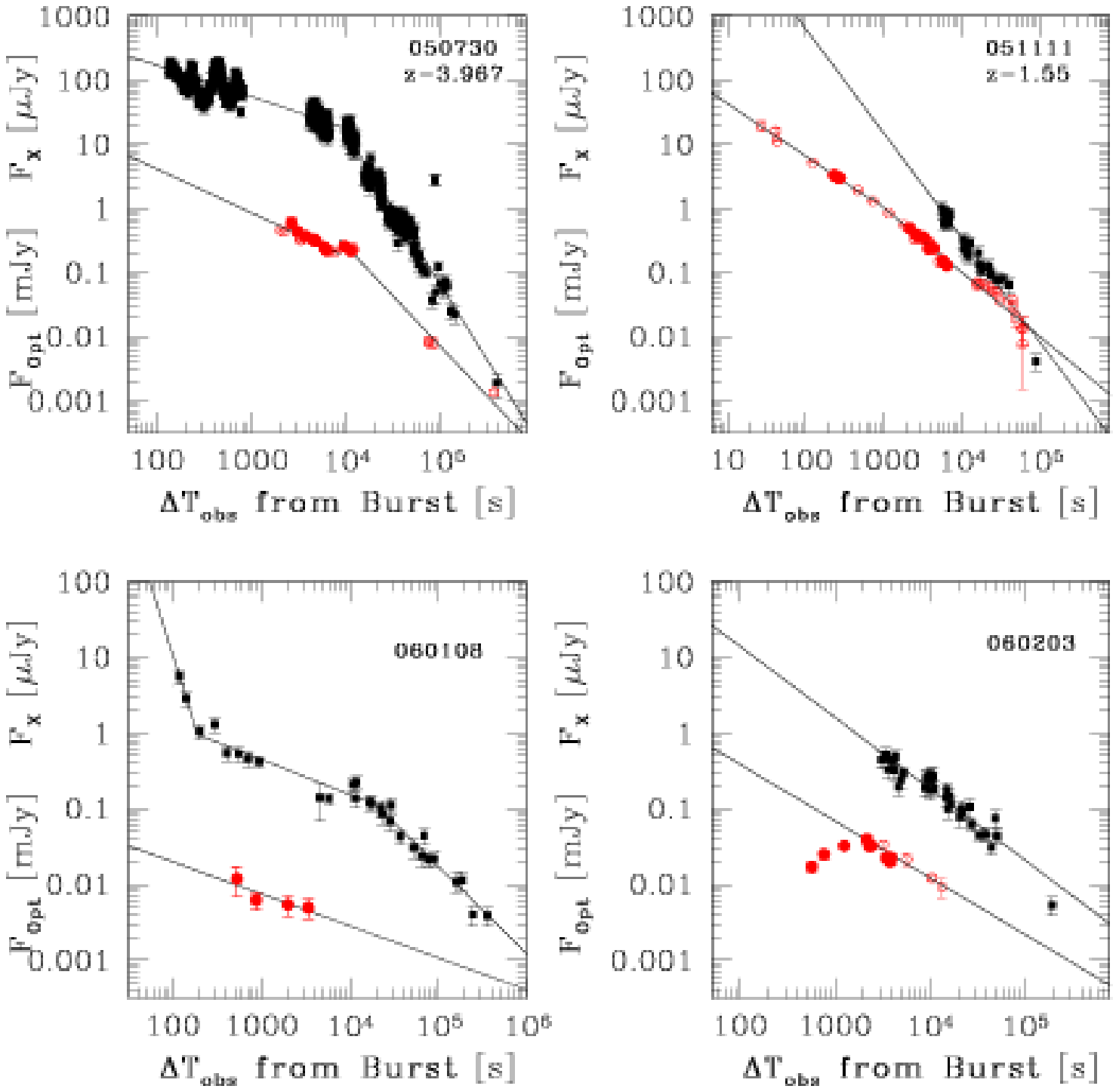}
\caption{\small X-ray and optical ($R$-band) light curves for the 
24 GRB afterglows detected by our telescopes from October 2004 to
September 2007. For each burst we show the X-ray flux density in
$\mu$Jy (black filled squares) and optical flux density in mJy (red
filled circles for our observations and red open circles for published
data, when available). We show also the value of the spectroscopic
redshift when available. X-ray data of {\it Swift}/XRT are from Evans
et al. (2007). We superimpose simple power-law fit segments to each
curve (the details of the fit are reported on Table 1).}
\label{figxopt} 
\end{figure}

\addtocounter{figure}{-1}

\begin{figure} \centering 
\includegraphics[width=15cm,height=9.5cm]{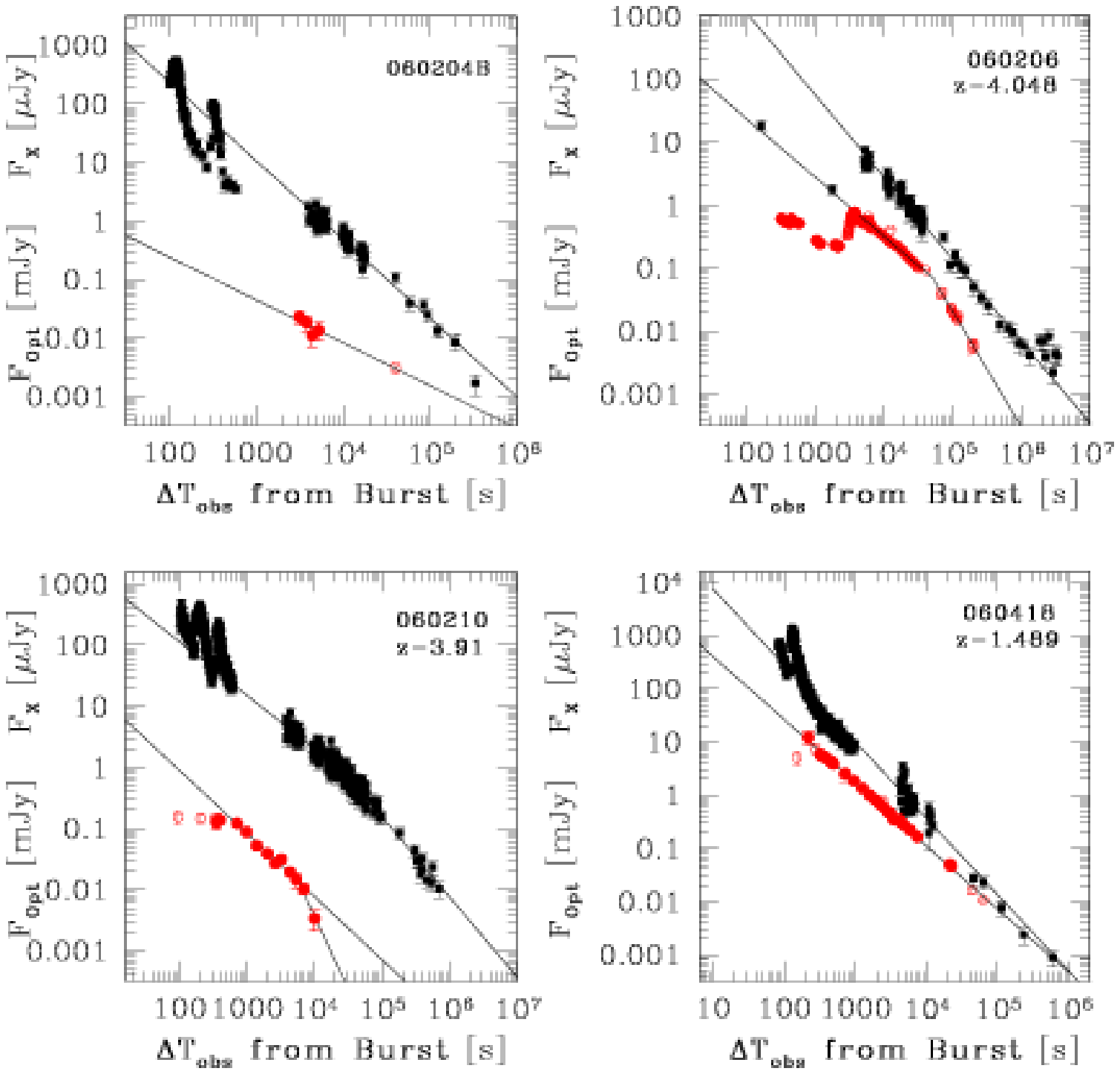}
\includegraphics[width=15cm,height=9.5cm]{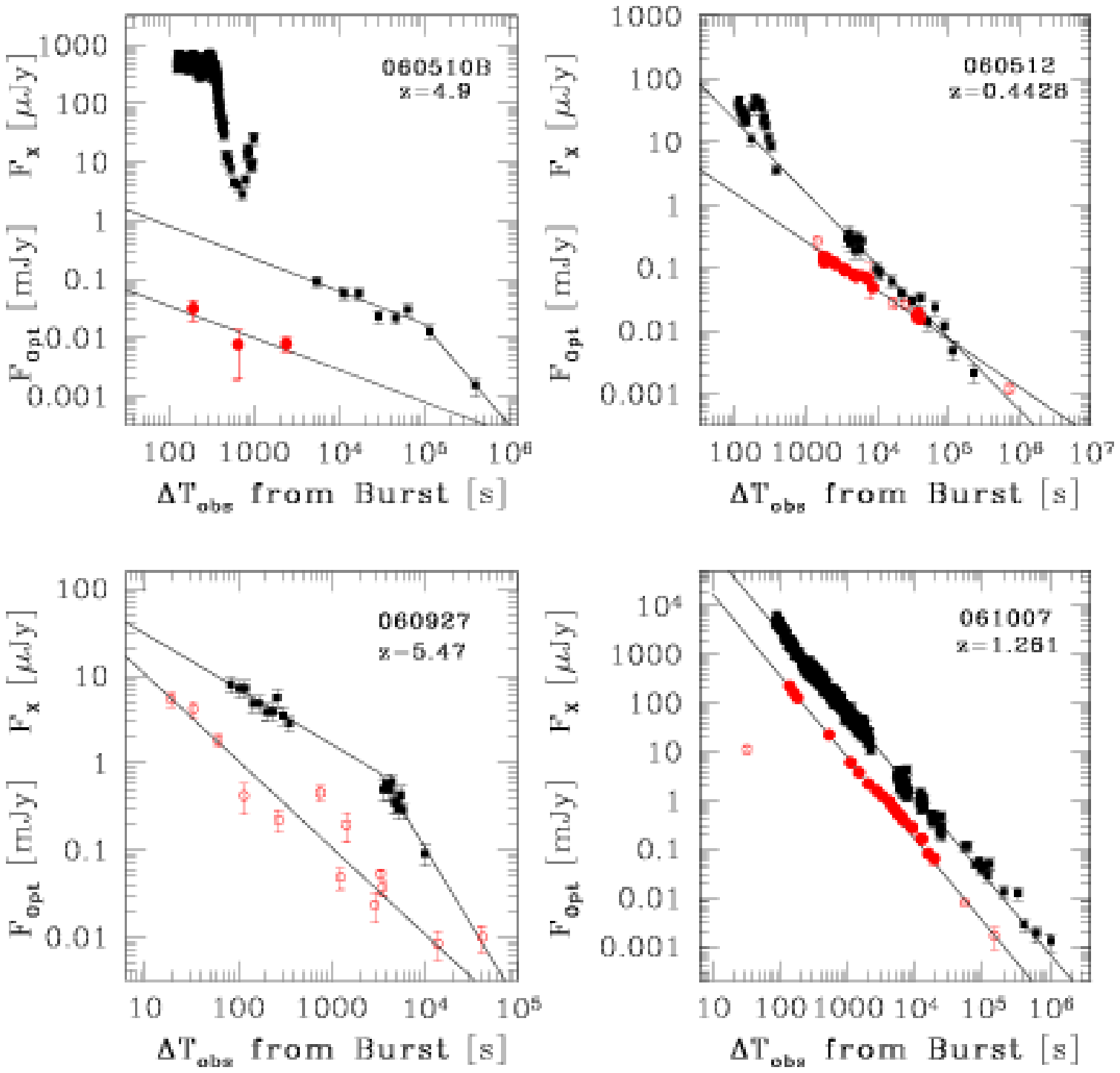}
\caption{- continued. For GRB~060927 there are no data in the 
R band because we detected this burst only in the $i'$ filter due to
its high redshift ($z=5.467$).}
\label{figxopt2} 
\end{figure}

\addtocounter{figure}{-1}

\begin{figure} \centering 
 \includegraphics[width=15cm,height=9.5cm]{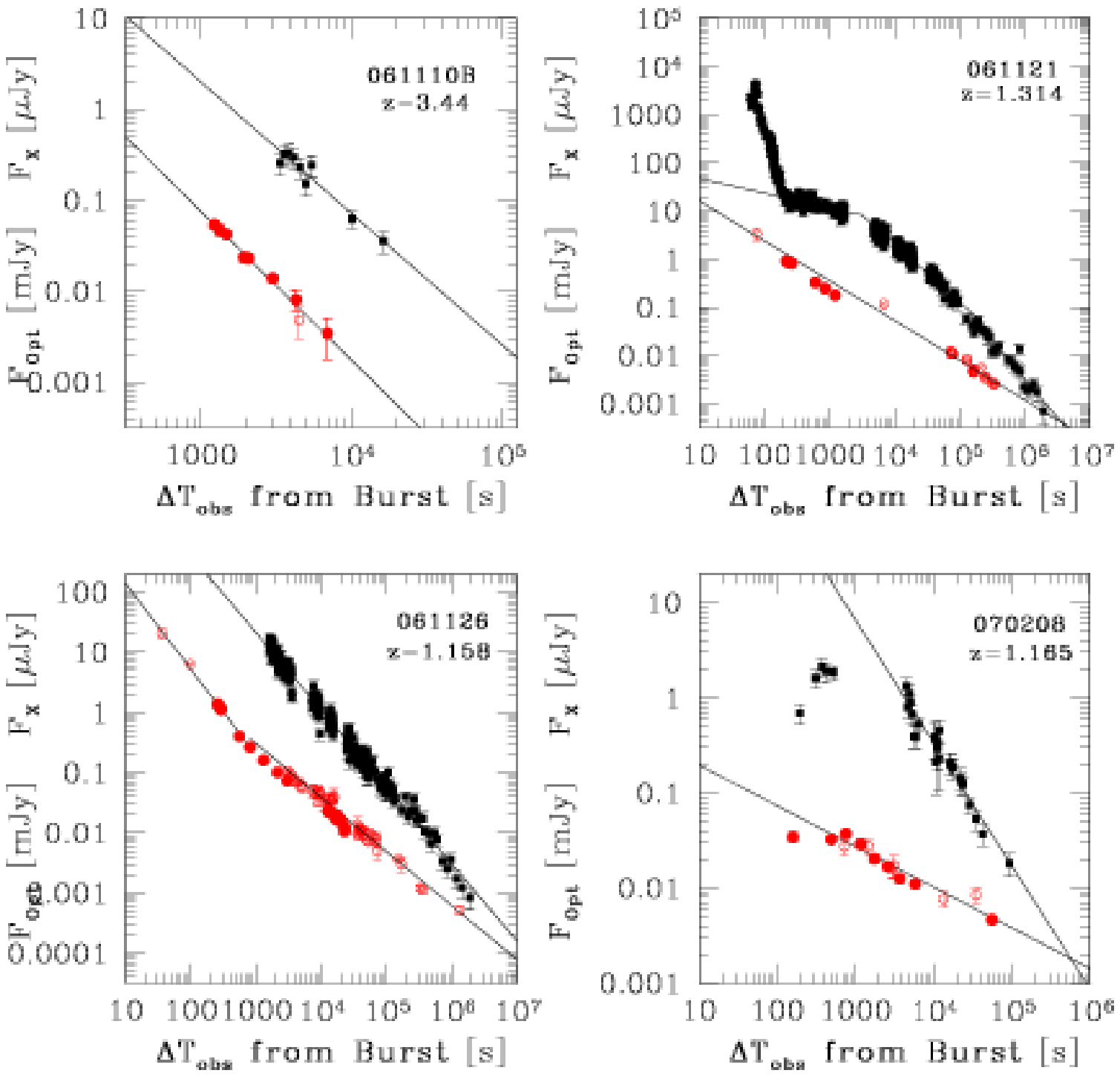}
 \includegraphics[width=15cm,height=9.5cm]{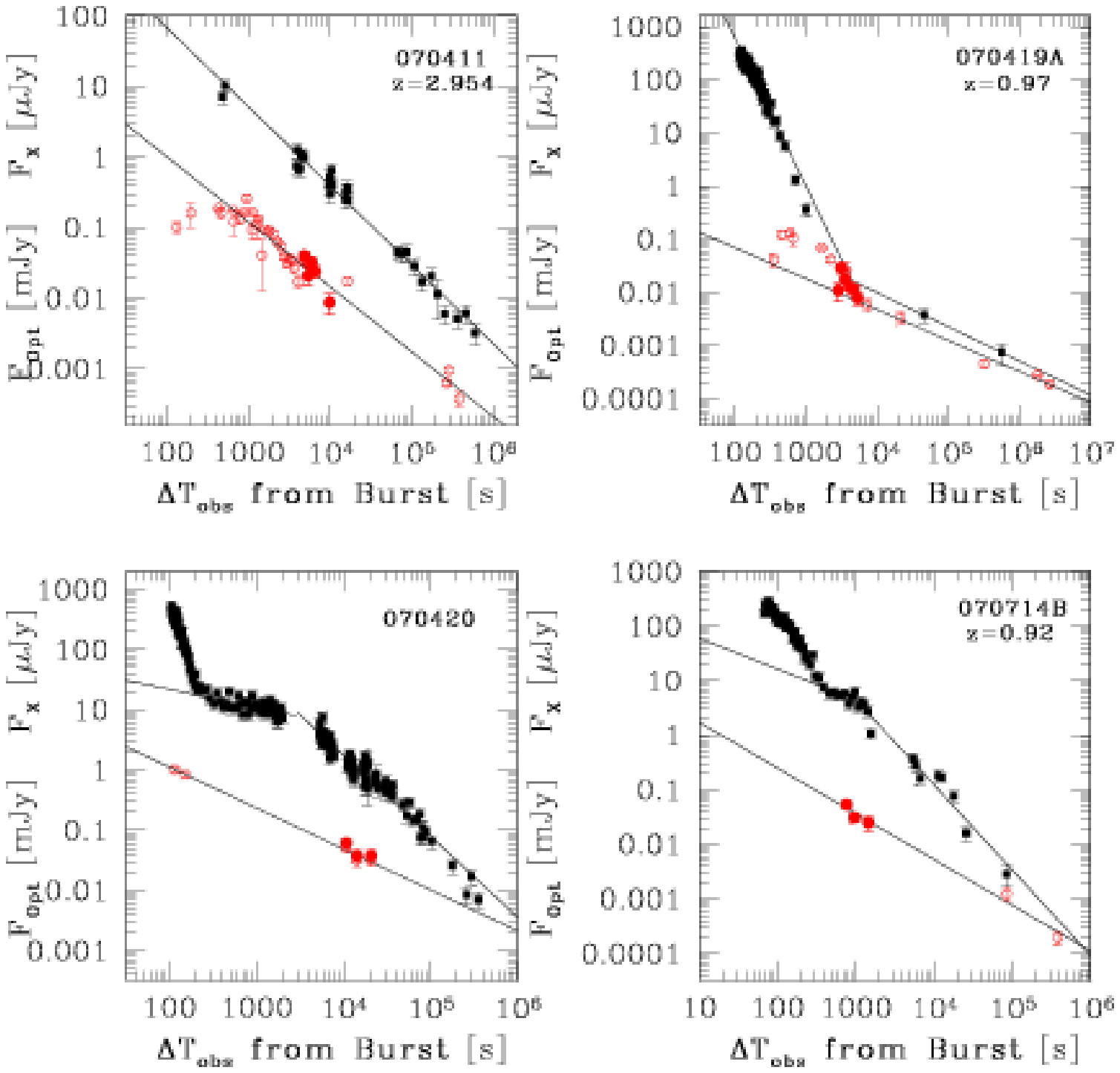}
\caption{- continued.}
\label{figxopt3} 
\end{figure}


\clearpage

\begin{figure*} \centering 
\includegraphics[width=15cm,height=10cm]{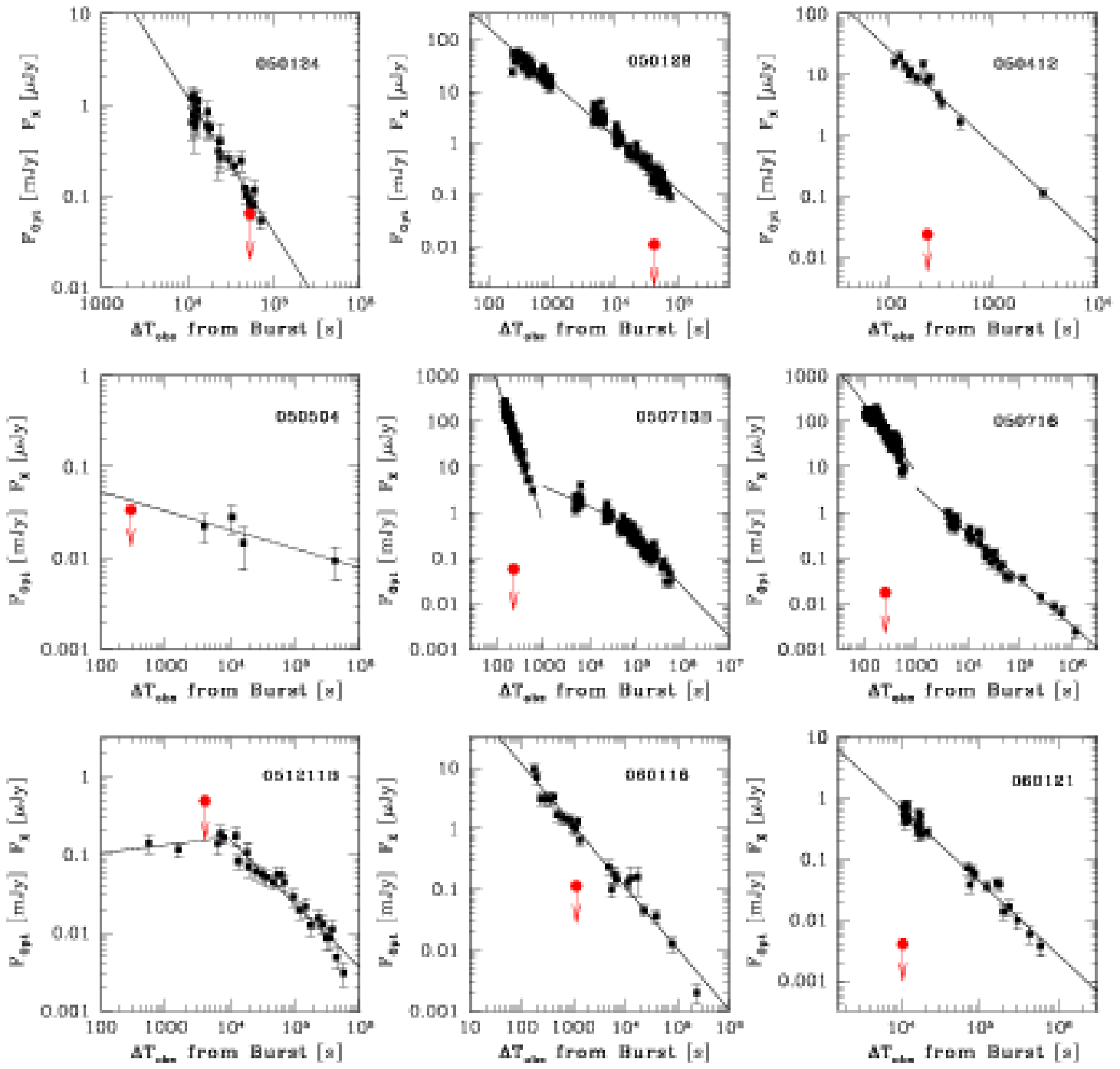}
\includegraphics[width=15cm,height=10cm]{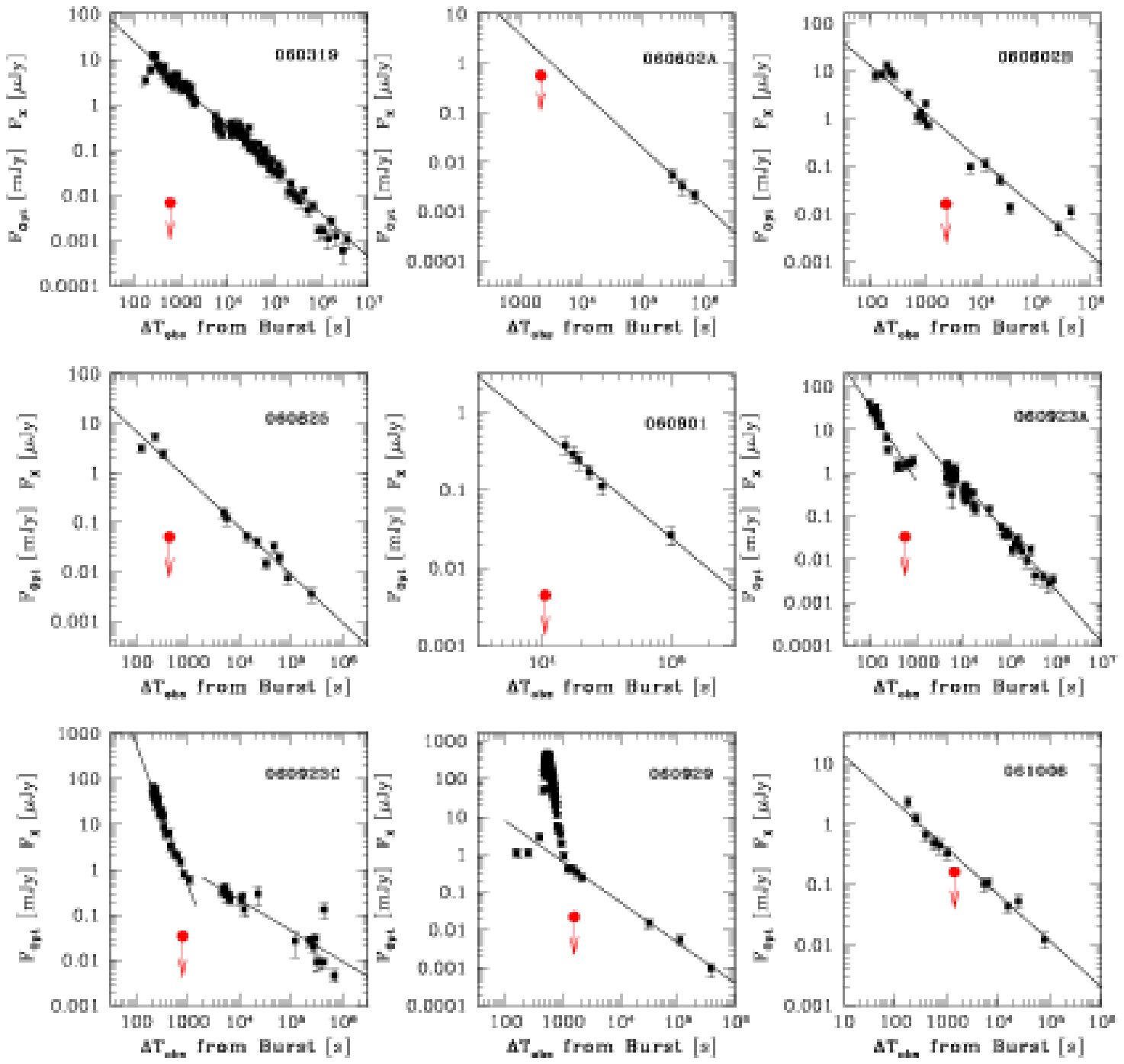}
\caption{X-ray light curves (black squares) for some GRBs for which 
we provide deep optical upper limit (red dots). These are the bursts
in Table 2 that have been observed by the {\it Swift}-XRT (data from
Evans et al. 2007).}
\label{figul2} 
\end{figure*}

\addtocounter{figure}{-1}

\begin{figure*} \centering 
\includegraphics[width=15cm,height=16.0cm]{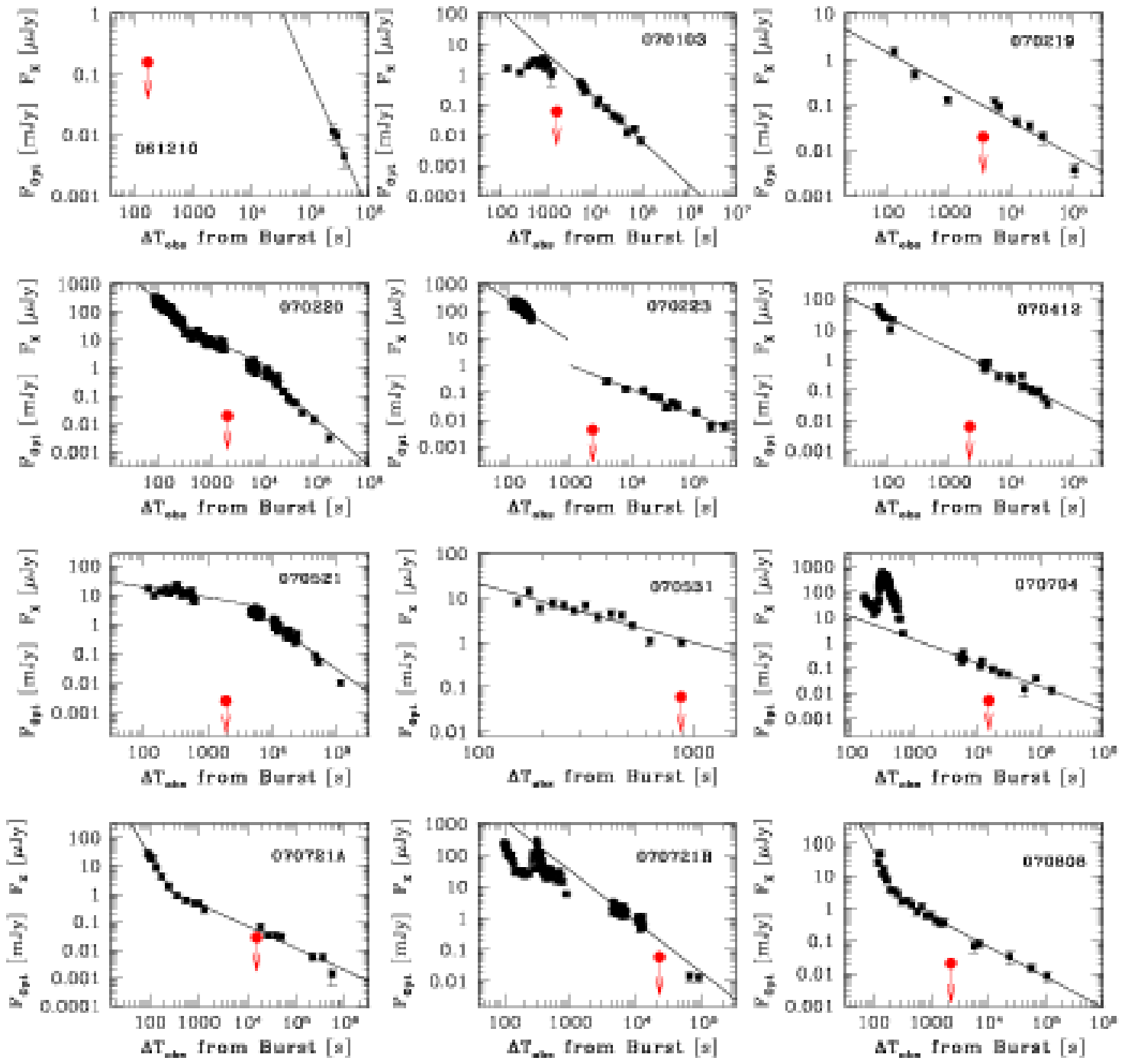}
\caption{- continued.}
\label{figul3} 
\end{figure*}


\clearpage

\begin{figure*} \centering 
\includegraphics[height=10cm,width=15cm]{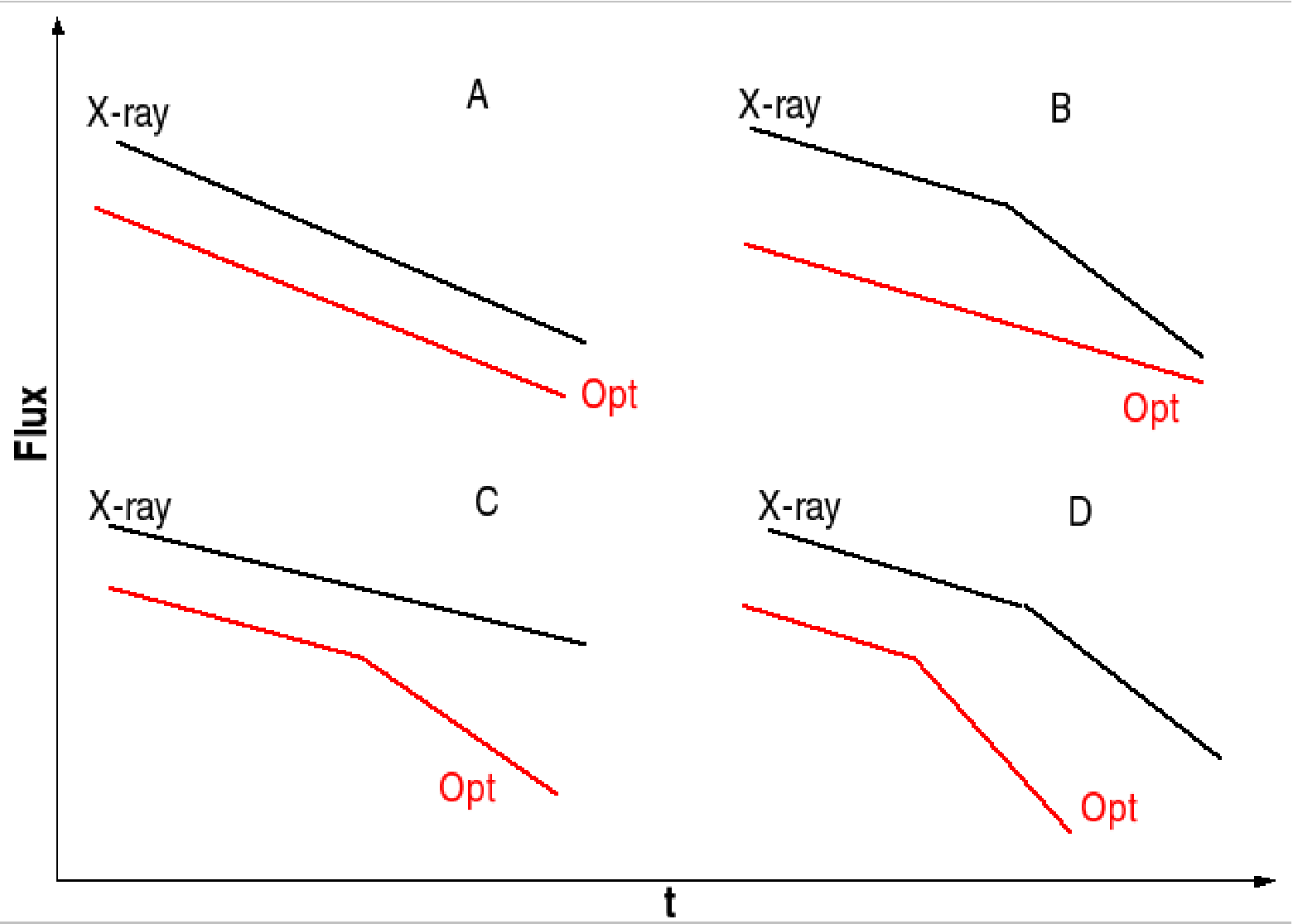}
\caption{Schematic view of the observed shapes of light curves in 
the optical (red) and X-ray band (black).}
\label{figc2} 
\end{figure*}


\clearpage

\begin{figure*} \centering 
\includegraphics[height=7cm,width=7cm]{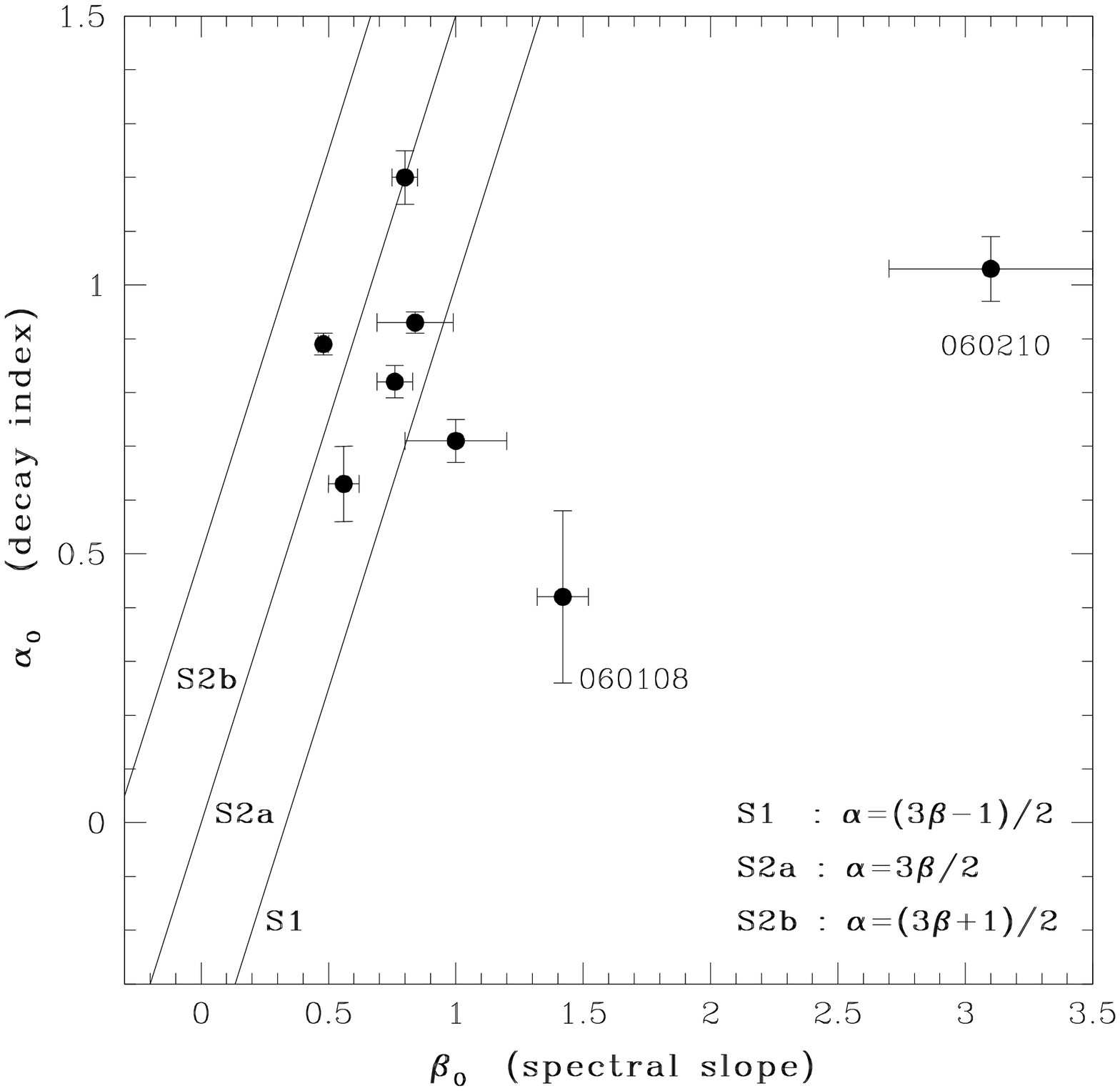}
\includegraphics[height=7cm,width=7cm]{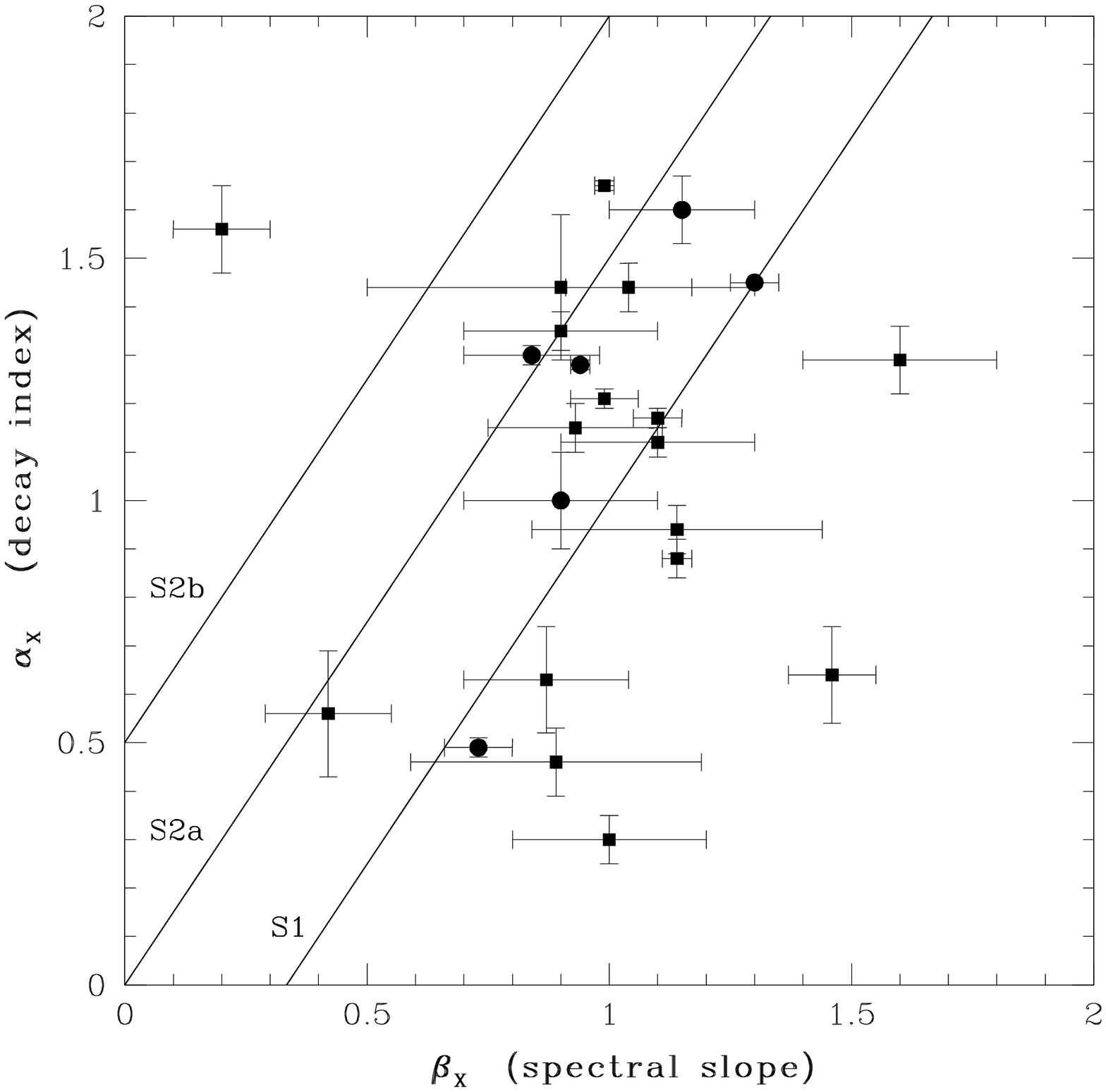}
\caption{Left panel : optical spectral slope ($\beta_{\rm O}$) vs optical 
temporal decay index ($\alpha_{\rm O}$). Right panel : spectral slope
($\beta_{\rm X}$) vs temporal decay ($\alpha_{\rm X}$) in the X-ray
band. The three lines drawn are the closure relations expected for the
standard fireball model: S1 = spherical outflow with the cooling
frequency ($\nu_{\rm c}$) below the observing frequency (optical or
X-ray), S2a = spherical outflow with $\nu_{\rm c}$ above the observing
frequency in a homogeneous medium, S2b = spherical outflow with
$\nu_{\rm c}$ above the observing frequency in a wind-like medium.}
\label{figab} 
\end{figure*}


\clearpage

\begin{figure*} \centering
\includegraphics[height=7cm,width=7cm]{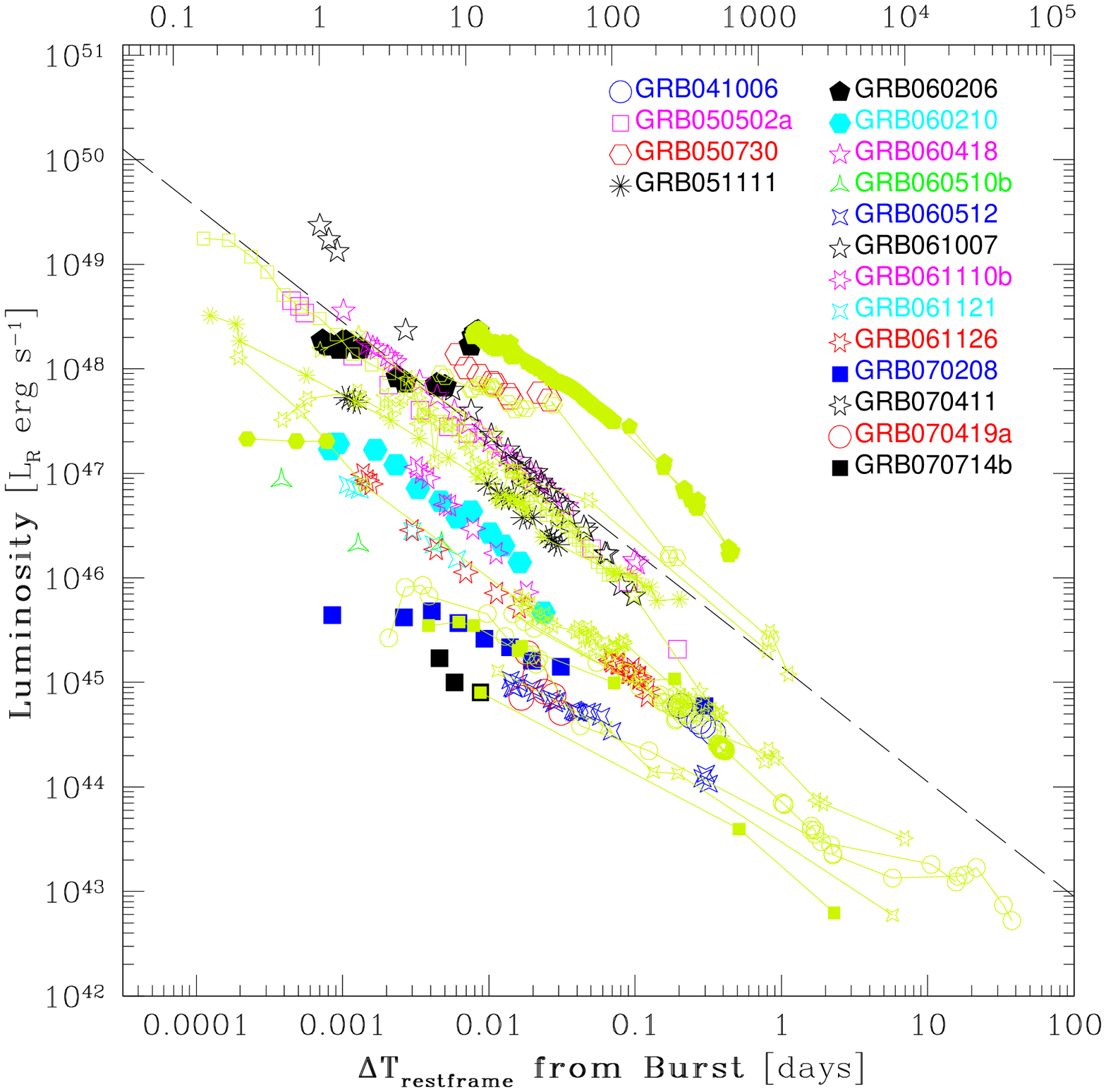}
\includegraphics[height=7cm,width=7cm]{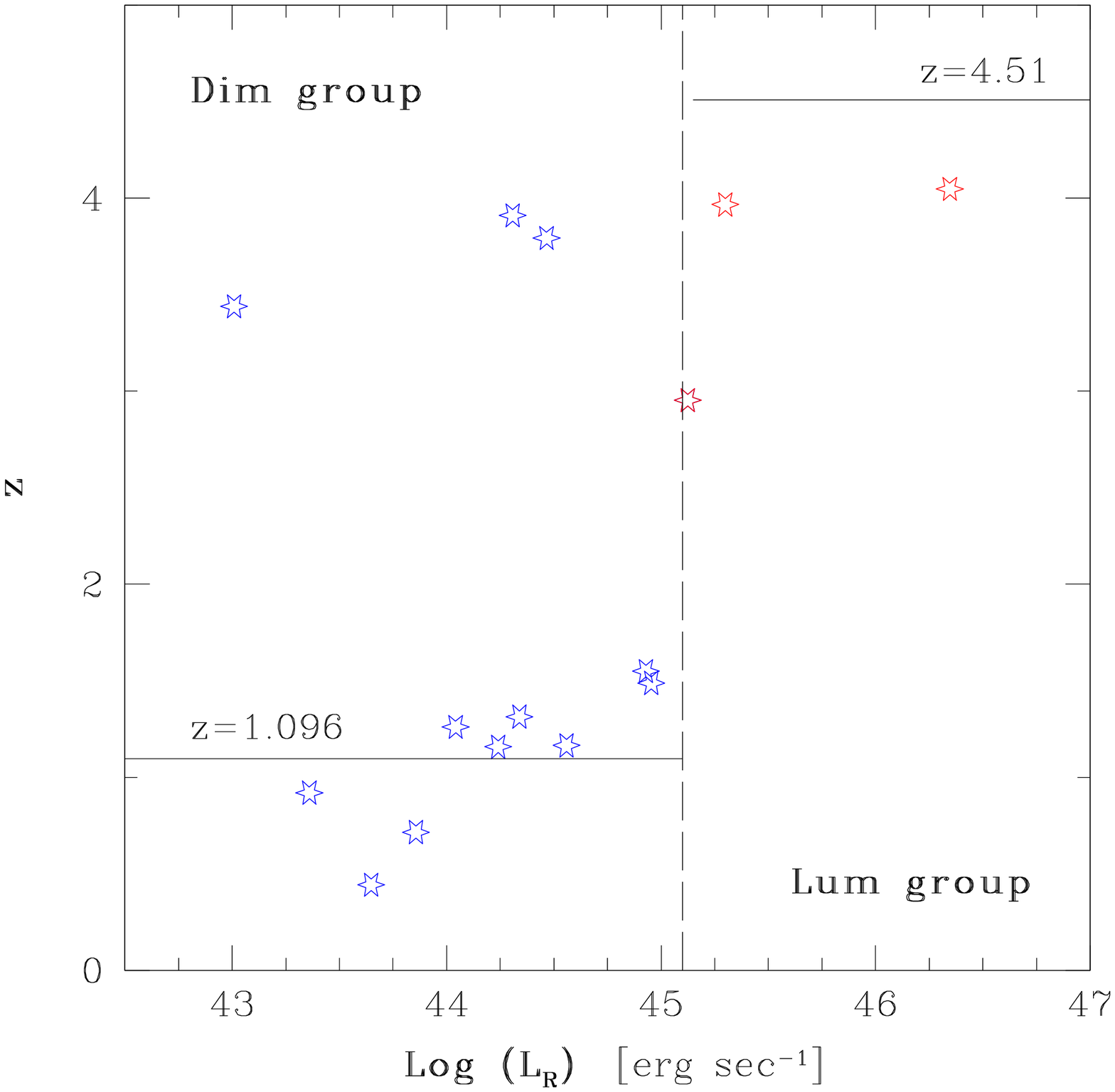} \caption{Left
panel: rest frame luminosity for all the afterglows in our sample with
known spectroscopic redshift. We superimposed on our data all the
published data (GCNs and refereed journal papers). The time axis is
given in days for an easier comparison with the similar plot of LZ06
and the same time in the restframe $\Delta T_{\rm rest-frame}$ is
given along the top of the plot in seconds to be consistent with the
earlier plots. The black dotted line shows the luminosity separation
($L_{*}$) between luminous and dim bursts as defined by LZ06, see text
for details. Right panel: luminosity rest frame at 1 day against
redshift for the bursts of our sample. The vertical line is $L_{*}$
and the two horizontal lines show the biggest values for the redshift
of the two classes of LZ06.}  \label{figall2}
\end{figure*}


\clearpage

\begin{figure} \centering 
\includegraphics[height=7cm,width=7cm]{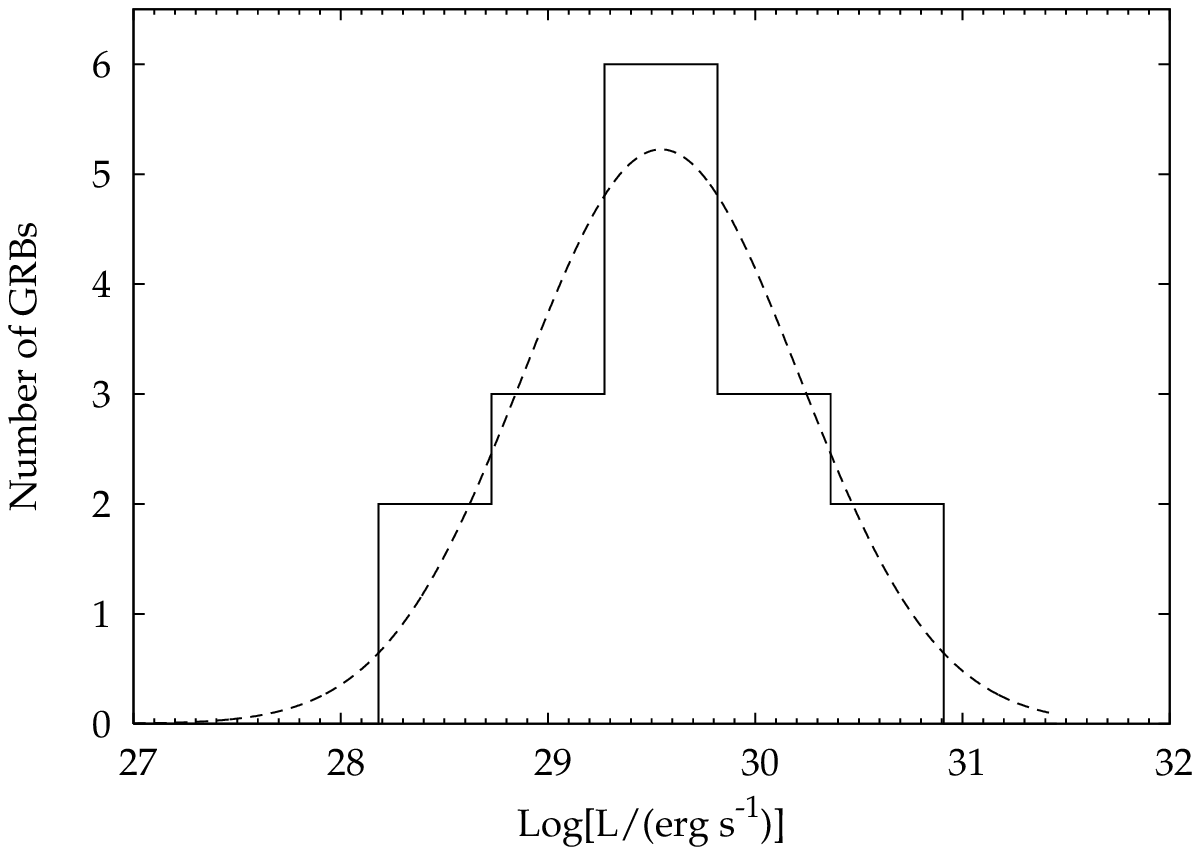}
\caption{The observed luminosity distribution of our sample 
at 12 hours, fitted with a single log-normal function with an average
of $29.54\pm0.07$ and a $\sigma$ of $0.67\pm0.05$.}
\label{figlum12h} 
\end{figure}


\clearpage

\begin{figure*} \centering 
\includegraphics[height=7cm,width=7cm]{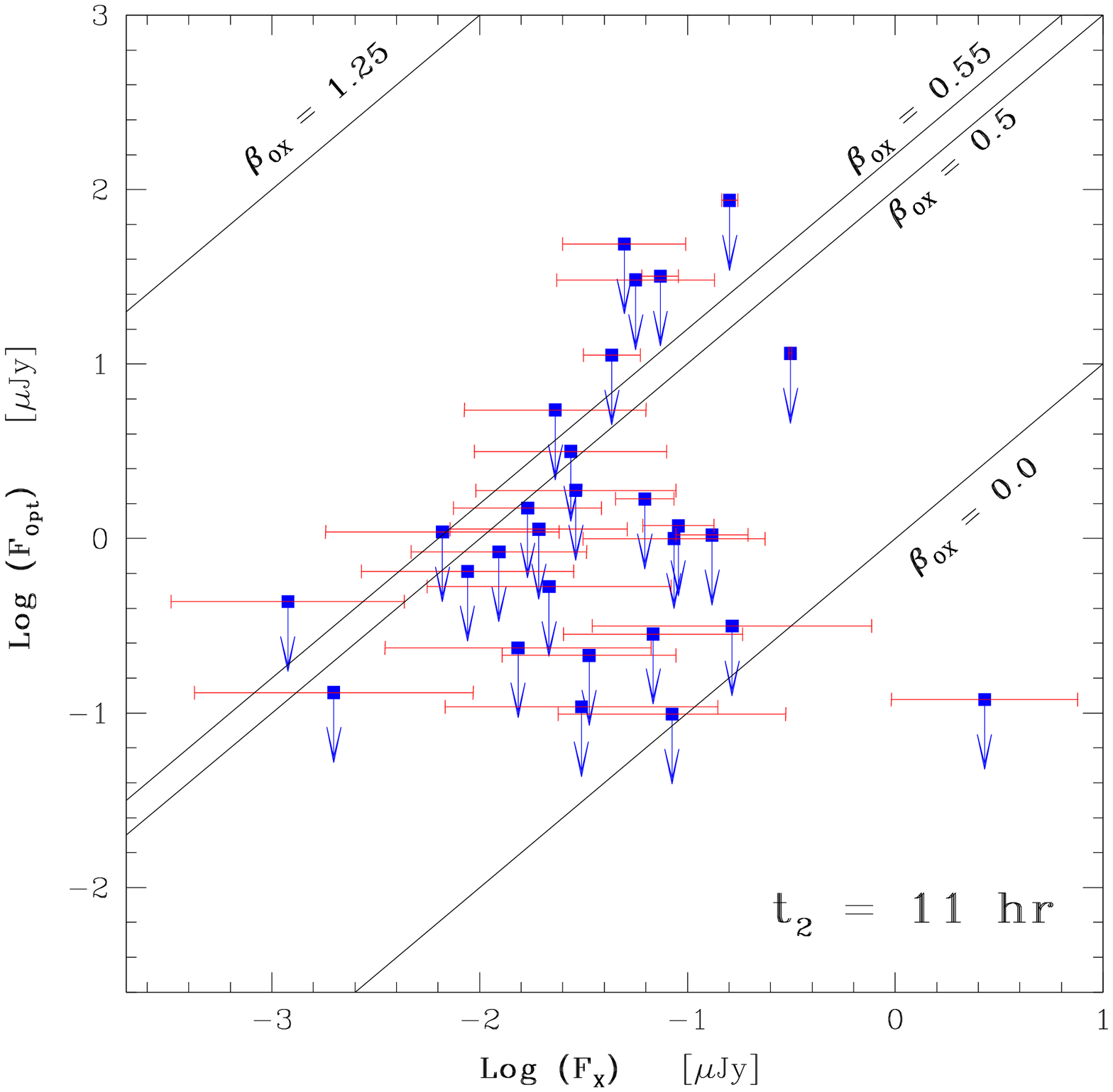}
\includegraphics[height=7cm,width=7cm]{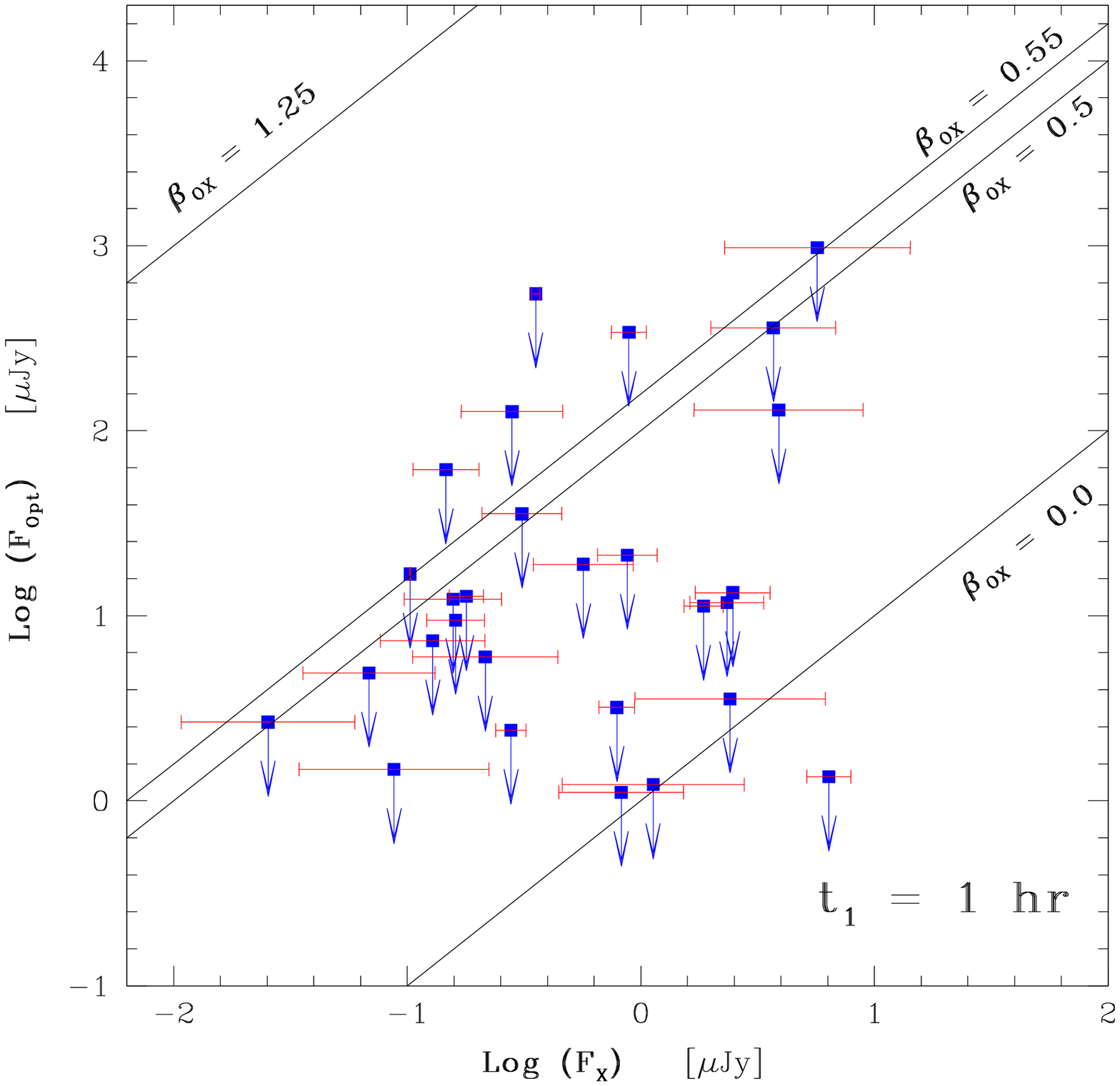}
\includegraphics[height=7cm,width=7cm]{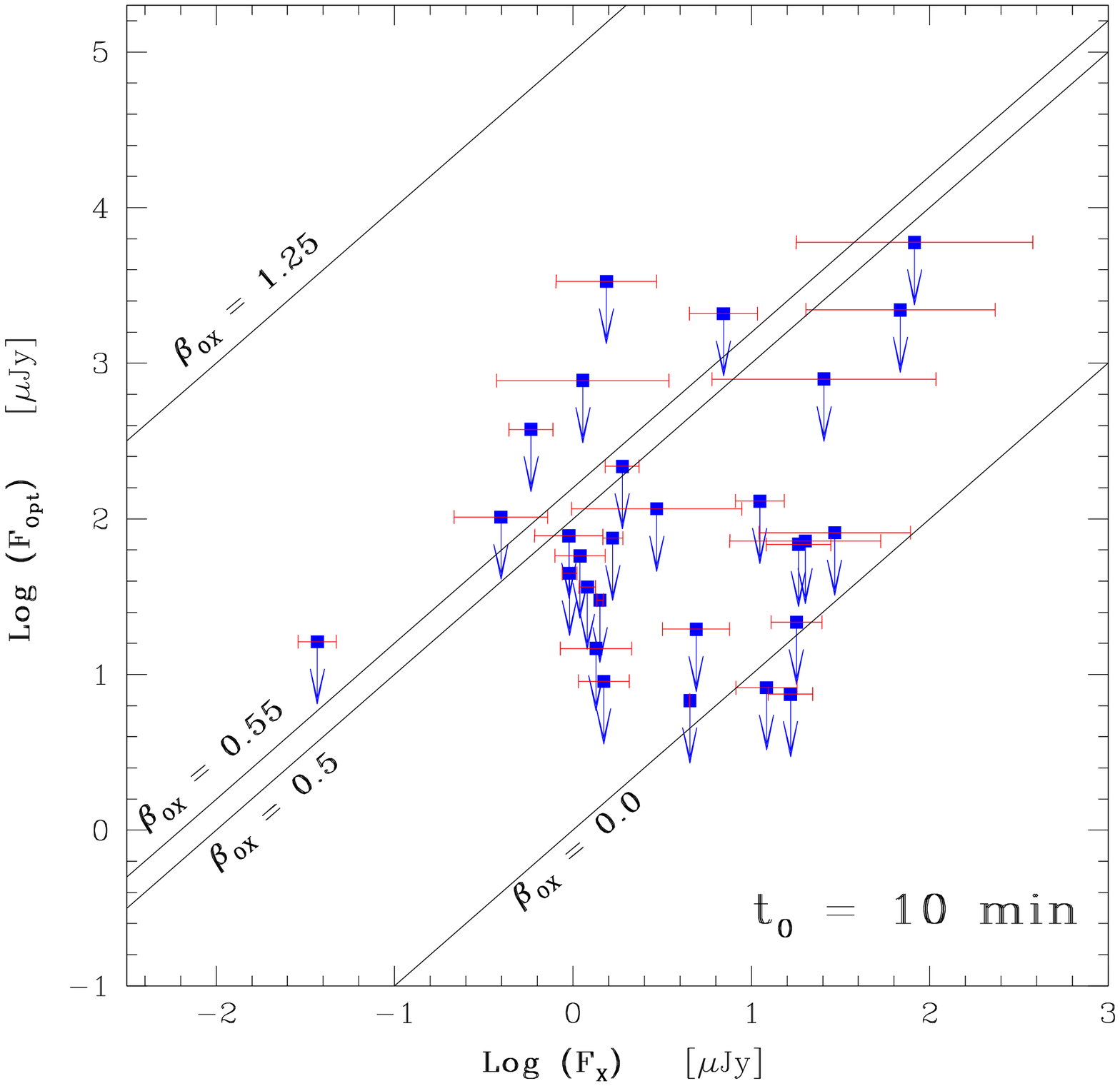}
\includegraphics[height=7cm,width=7cm]{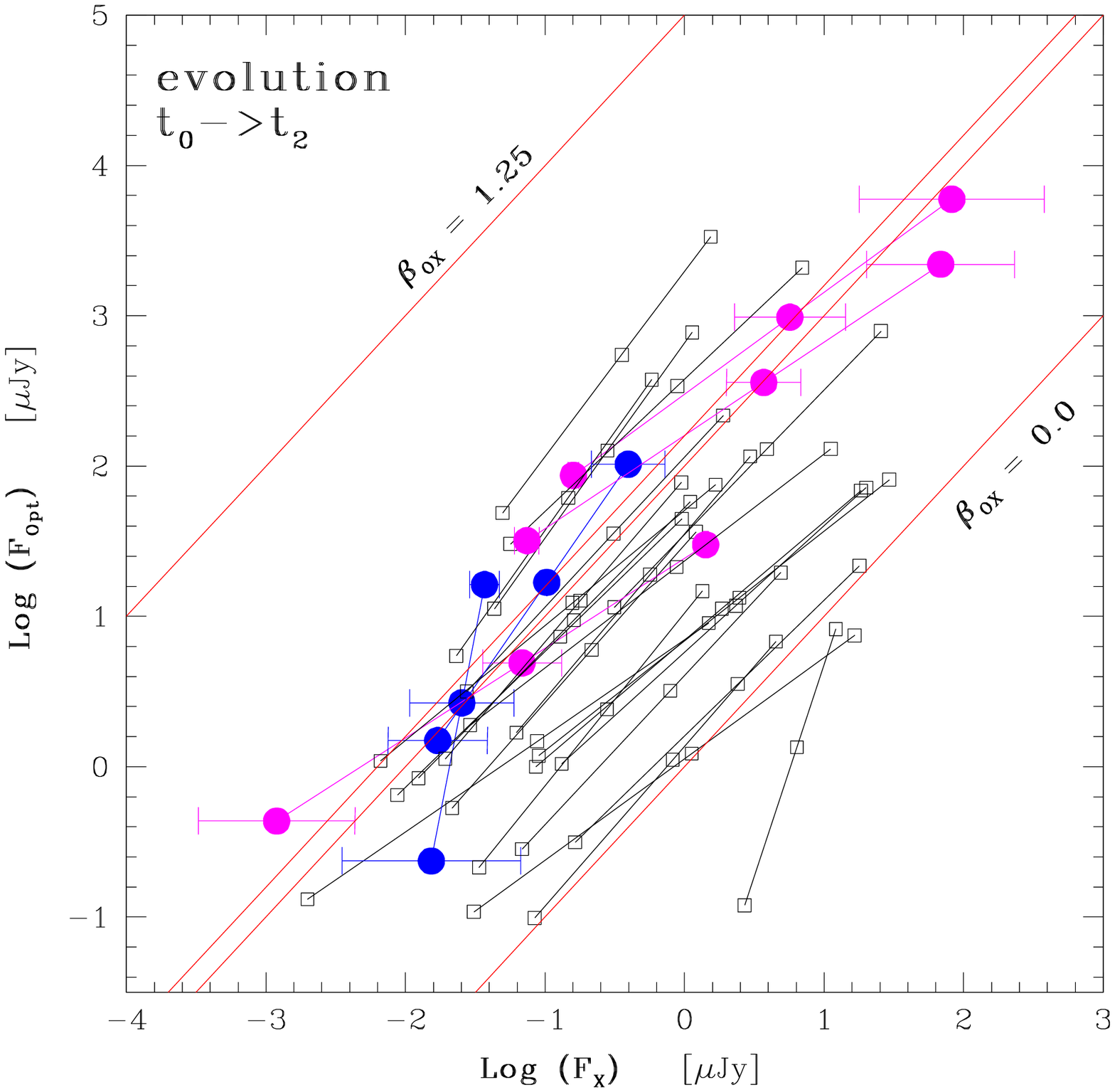}
\caption{Plots of optical flux ($F_{\rm O}$) versus X-ray flux ($F_{\rm X}$) 
for all the bursts listed in Table~2 for which an XRT observation was
available. Extrapolation of the fluxes has been done at $t_{\rm 2}$=11
hours (top left), $t_{\rm 1}$=1 hour (top right) and $t_{0}$=10
minutes (bottom left). Lines with constant $\beta_{\rm OX}$ are
shown. Dark bursts are the ones below the constant line $\beta_{\rm
OX}=0.5$. The plot at bottom right shows the evolution of the optical
and X-ray flux from $t_{\rm 0}$ to $t_{\rm 2}$; for clarity the errors
on this panel are not shown.}
\label{figuljako} 
\end{figure*}


\clearpage

\setcounter{table}{0}
\begin{landscape}
\begin{deluxetable}{@{}cccccccccccccc}
\tabletypesize{\tiny}
\tablewidth{0pt}
 \tablecaption{\tiny Optical and X-ray light curves parameters. If data are well fitted with a broken power law then $\alpha_{1}$ and $\alpha_{2}$ represent the decay index pre- and post-break respectively and $t_{\rm break}$ is the break time, for both optical and X-ray bands. The value of the reduced $\chi^{2}$ ($\chi^{2}/\nu$, where $\nu$ are the degrees of freedom) is given for each fit in the two bands. The values of $\beta_{\rm O}$, $\beta_{\rm X}$ and $\beta_{\rm OX}$ are the slopes of the spectral energy distribution taken from the literature. In the last two columns we report the value of the redshift ($z$) and the isotropic energy ($E_{\gamma, {\rm iso}}$) of the burst when available. * = this value refers to $R$ band data taken from GCNs because we detected GRB~060927 only in the $i'$ band, due to the high redshift of the event.}
\label{tabsample}
\tablewidth{0pt}
\tablehead{
\colhead{GRB} & \colhead{$\alpha_{{\rm O},1}$} & \colhead{$\alpha_{{\rm O},2}$} & \colhead{$t_{{\rm O},break}$} & \colhead{$\chi^{2}/\nu$} & \colhead{$\alpha_{{\rm X},1}$} & \colhead{$\alpha_{{\rm X},2}$} & \colhead{$t_{{\rm X},break}$} & \colhead{$\chi^{2}/\nu$} & \colhead{$\beta_O$} & \colhead{$\beta_X$} & \colhead{$\beta_{OX}$} & \colhead{$z$} & \colhead{$E_{\gamma, {\rm iso}}$}\\
\colhead{} & \colhead{} & \colhead{} & \colhead{(days)} & \colhead{} & \colhead{} & \colhead{} & \colhead{(days)} & \colhead{} & \colhead{} & \colhead{} & \colhead{} & \colhead{} & \colhead{[$10^{52}$ erg]}}
\startdata
041006 & $0.71\pm0.04$ & $1.24\pm0.01$ & $0.15\pm0.01$ & $77/72$ &--- & $1.0\pm0.1$ & --- & --- & $1.0\pm0.2$ & $0.9\pm0.2$ & $\sim0.7$ & 0.716 & $0.94_{-0.08}^{+0.21}$ \\
041218 & $1.25\pm0.10$ & $1.47\pm0.13$ & $0.11\pm0.01$ & $19/16$ & --- & --- & --- & --- & --- & --- & --- & --- & ---\\
050502A & $1.20\pm0.04$ & --- & --- & $50/52$ & $>1.45$ & --- & --- & --- & $0.80\pm0.05$ & $1.30\pm0.05$ & $0.8\pm0.1$ & 3.793 & $4_{-1}^{+3}$ \\
050713A & $0.63\pm0.04$ & --- & --- & $12/6$ & $1.17\pm0.05$ & $1.32\pm0.09$ & $0.28\pm0.06$ & $53/45$ & --- & $1.10\pm0.05$ & $1.2\pm0.1$ & --- & --- \\
050730 & $0.63\pm0.07$ & $1.55\pm0.08$ & $0.10\pm0.01$ & $46/45$ & $0.49\pm0.02$ & $2.37\pm0.05$ & $0.11\pm0.01$ & $670/566$ & $0.56\pm0.06$ & $0.73\pm0.07$ & --- & 3.967 & $9_{-3}^{+8}$ \\
051111 & $0.82\pm0.03$ & $1.00\pm0.03$ & $0.008\pm0.001$ & $46/43$ & --- & $1.60\pm0.07$ & --- & $34/29$ & $0.76\pm0.07$ & $1.15\pm0.15$ & $0.84\pm0.02$ & 1.55 & $6_{-2}^{+5}$ \\
060108 & $0.42\pm0.16$ & --- & --- & $5/4$ & $0.46\pm0.07$ & $1.15\pm0.07$ & $0.18\pm0.02$ & $29/25$ & $\sim1.4$ & $0.89\pm0.30$  & $0.54\pm0.10$ & $<3.2$ & $<0.795$\\
060203 & $0.74\pm0.13$ & --- & --- & $11/8$ & $0.94\pm0.05$ & --- & --- & $41/34$ & --- & $1.24\pm0.30$ & --- & --- & --- \\
060204B & $0.73\pm0.10$ & --- & --- & $4/3$ & --- & $1.35\pm0.04$ & --- & $54/60$ & --- & $0.9\pm0.2$ & --- & --- & ---\\
060206 & $0.93\pm0.02$ & $1.83\pm0.02$ & $0.60\pm0.01$ & $235/216$ & $1.30\pm0.02$ & --- & --- & $66/72$ & $0.84\pm0.15$ & $0.84\pm0.14$ & $0.93\pm0.02$ & 4.048 & $4.1_{-0.6}^{+1.2}$ \\
060210 & $1.03\pm0.06$ & ($2.38\pm0.43$) & ($0.10\pm0.02$) & $9/7$ & $0.88\pm0.04$ & $1.31\pm0.03$ & $0.33\pm0.03$ & $263/221$& $3.1\pm0.4$ & $1.14\pm0.03$ & $0.3\pm0.1$ & 3.91 & $42_{-8}^{+35}$\\
060418 & $1.19\pm0.02$ & --- & --- & $29/26$ & $1.44\pm0.05$ & --- & --- & $97/80$ & --- & $1.04\pm0.13$ & --- & 1.489 & $10_{-2}^{+7}$\\
060510B & $0.55\pm0.34$ & --- & --- & $2/1$ & $0.56\pm0.13$ & $1.7\pm0.1$ & $1.16\pm0.02$ & $5/6$ & --- & $0.42\pm0.13$ & --- & 4.9 & $23_{-4}^{+10}$\\
060512 & $0.77\pm0.02$ & --- & --- & $19/22$ & $1.15\pm0.05$ & --- & --- & $22/17$ &  --- & $0.93\pm0.18$ & --- & 0.4428 & $0.020_{-0.004}^{+0.030}$\\
060927 & $0.99\pm0.11^{*}$ & --- & --- & $19/11$ & $0.63\pm0.11$ & $1.78\pm0.21$ & $0.035\pm0.003$ & $16/13$ & ---& $0.87\pm0.17$ & --- & 5.467 & $9_{-1}^{+2}$ \\
061007 & $1.71\pm0.02$ & --- & --- & $24/23$ & $1.65\pm0.01$ & --- & --- & $954/978$ & --- & $0.99\pm0.02$ & $1.02\pm0.05$ & 1.261 & $140_{-60}^{+110}$\\
061110B & $1.64\pm0.08$ & --- & --- & $8/7$ & $1.44\pm0.15$ & --- & --- & $11/7$ & --- & $0.9\pm0.4$ & --- & 3.44 & $13_{-6}^{+16}$\\
061121 & $0.83\pm0.03$ & --- & --- & $41/36$ & $1.21\pm0.02$ & $1.58\pm0.12$ & $2.89\pm0.03$ & $200/191$ & ---- & $0.99\pm0.07$ & $0.53\pm0.06$ & 1.314 & $19_{-5}^{+11}$\\
061126 & $1.43\pm0.12$ & $0.89\pm0.02$ & $0.009\pm0.001$ & $93/72$ & --- & $1.28\pm0.01$ & --- & $348/273$ & $0.48\pm0.02$ & $0.98\pm0.02$ & $0.53\pm0.02$ & 1.158 & $8_{-2}^{+7}$\\
070208 & $0.42\pm0.04$ & --- & --- & $14/11$ & --- & $1.29\pm0.07$ & --- & $36/24$ & --- & $1.6\pm0.2$ & --- & 1.165 & $0.28_{-0.08}^{+0.22}$ \\
070411 & $0.92\pm0.04$ & --- & --- & $61/38$ & $1.12\pm0.03$ & --- & --- & $39/28$ & --- & $1.1\pm0.2$ & --- & 2.954 &  $10_{-2}^{+8}$\\
070419A & --- & $0.58\pm0.04$ & --- & $21/9$ & $2.79\pm0.06$ & $0.64\pm0.10$ & $0.046\pm0.005$ & $139/105$ & --- & $1.46\pm0.09$ & --- & 0.97 & $0.24_{-0.03}^{+0.23}$ \\
070420 & $0.68\pm0.03$ & --- & --- & $4/3$ & $0.30\pm0.05$ & $1.34\pm0.03$ & $0.034\pm0.04$ & $232/142$ & --- & $1.0\pm0.2$ & --- & --- & --- \\
070714B & $0.83\pm0.04$ & --- & --- & $5/3$ & $0.56\pm0.19$ & $1.56\pm0.09$ & $0.010\pm0.02$ & $28/19$ & --- & $0.2\pm0.1$ & --- & 0.92 & $0.8_{-0.1}^{+2.0}$\\
\enddata
\tablecomments{Notes - References for $\alpha_{{\rm X},1}$ : GRB~050502A : \cite{gui1b}. References for $\alpha_{{\rm X},2}$ : GRB~041006 : \cite{butler2}. References for $\beta_{\rm O}$ : GRB~041006 : \cite{garna}; GRB~050502A : \cite{gui1b,yost}; GRB~050730 : \cite{pandey}; GRB~051111 : \cite{guidorzi2}; GRB~060108 : \cite{oates}; GRB~060206 : \cite{monfardini}; GRB~060210 : \cite{curran}; GRB~061126 : \cite{gomb2}. References for $\beta_{\rm X}$ : GRB~041006 : \cite{butler2}; GRB~050502A : \cite{gui1b}; GRB~050713A : \cite{morris}; GRB~050730 : \cite{pandey}; GRB~051111 : \cite{guidorzi2}; GRB~060108 : \cite{oates}; GRB~060203 : \cite{mor1}; GRB~060204B : \cite{falcone}; GRB~060206 : \cite{monfardini}; GRB~060210 : \cite{curran}; GRB~060418 : \cite{falcone2}; GRB~060510B : \cite{perri}; GRB~060512 : \cite{godet}; GRB~060927 : \cite{ruiz}; GRB~061007 : \cite{schady}; GRB~061110B : \cite{grupe}; GRB~061121 : \cite{page}; GRB~061126 : \cite{gomb2}; GRB~070208 : \cite{conc1}; GRB~070411 : \cite{moretti}; GRB~070419A : \cite{perri2}; GRB~070420 : \cite{stratta}; GRB~070714B : \cite{racusin}. References for $\beta_{\rm OX}$ : GRB~041006 : \cite{butler2}. GRB~050502A : \cite{gui1b}; GRB~050713A : \cite{morris}; GRB~051111 : \cite{guidorzi2}; GRB~060108 : \cite{oates}; GRB~060206 : \cite{monfardini}; GRB~060210 : \cite{curran}; GRB~061007 : \cite{mundell2}; GRB~061121 : \cite{page}; GRB~061126 : \cite{gomb2}. References for $z$ : GRB~041006 : \cite{fugazza}; GRB~050502A : \cite{prochaska}; GRB~050730 : \cite{chen}; GRB~051111 : \cite{hill}; GRB~060108 : \cite{oates}; GRB~060206 : \cite{prochaska2}; GRB~060210 : \cite{cucchiara}; GRB~060418 : \cite{dupree,wrees}; GRB~060510B : \cite{price2}; GRB~060512 : \cite{bloom}; GRB~060927 : \cite{ruiz}; GRB~061007 : \cite{osip}; GRB~061110B : \cite{fynbo3}; GRB~061121 : \cite{bloom2}; GRB~061126 : \cite{perley}; GRB~070208 : \cite{cucchiara2}; GRB~070411 : \cite{jako2}; GRB~070419A : \cite{cenko}; GRB~070714B : \cite{graham}. References for $E_{\gamma, {\rm iso}}$ : \citep[all values from][ except for GRB~041006, GRB~050502A and GRB~060108]{butler4}.}

\end{deluxetable}
\end{landscape}


\clearpage

\setcounter{table}{1}

\begin{landscape}
\begin{deluxetable}{@{}ccccccccrrrrrcccc}
\tabletypesize{\tiny}
\tablewidth{0pt}
 \tablecaption{\scriptsize Upper limit parameters. Refer to Section 3.2 for
 detailed explanation of the colums of that table.}
\label{tabul}
\tablehead{\colhead{GRB} & \colhead{XRT} & \colhead{Duration} & \colhead{$\it f$ $\times 10^{7}$} & \colhead{F$_{\rm X}$ $\times 10^{11}$} & \colhead{$\Delta~t_{\rm X}$} & \colhead{$\alpha_{\rm X}$} & \colhead{$\beta_{\rm X}$} & \colhead{$\Delta~t_{\rm start}$} & \colhead{$R^{\rm u.l.}_{\rm start}$} & \colhead{$\Delta~t_{\rm mean}$} & \colhead{$R^{\rm u.l.}_{\rm mean}$} & \colhead{$T_{exp}$} & \colhead{OT} & \colhead{$A_{R}$} & \colhead{$\alpha_{\rm X}^{\rm (fit)}$} & \colhead{F$_{\rm X}$ ($\Delta~t_{R}$)}\\
\colhead{} & \colhead{pos.} & \colhead{(s)} & \colhead{(erg~cm$^{-2}$)} & \colhead{(erg~cm$^{-2}$~s$^{-1}$)} & \colhead{(min)} & \colhead{} & \colhead{} & \colhead{(min)} &\colhead{} & \colhead{(min)} & \colhead{} & \colhead{(min)} &\colhead{} & \colhead{} & \colhead{} & \colhead{($\mu$~Jy)}}
\startdata
041211 & no & 30.2 & 100 & --- & --- & --- & --- & 197.20 & 19.18 & 242.94 & 20.86 & 4.5 & --- & 0.45 & --- & --- \\ 
050124 & yes & 4.1 & 12.3 & 6.9 & 185.2 & --- & 0.3 & 885.64 & 19.00 & 886.02 & 19.20 & 0.5 & IR & 0.09 & $1.49\pm0.08$ & 0.103\\
050128 & yes & 13.8 & 51.7 & 24.0 & 3.62 & --- & --- & 697.20 & 21.13 & 788.4 & 21.85 & 20.0 & --- & 0.21 & $1.05\pm0.02$ & 0.296\\
050412 & yes & 26.0 & 5.66 & 0.39 & 166.7 & 1.35 & 0.4 & 2.5 & 18.7 & 3.83 & 20.82 & 11.5 & --- & 0.05 & $1.58\pm0.09$ & 6.790\\
050504 & yes & 80.0 & 15.0 & --- & 326.8 & --- & --- & 3.7 & 19.00 & 17.39 & 20.33 & 17.67 & --- & 0.03 & $0.21\pm0.08$ & 0.043\\
050520 & yes & 80.0 & 24.0 & 0.01 & 127.7 & 1.4 & --- & 4.5 & 16.60 & 8.34 & 19.4 & 2.5 & --- & 0.04 & --- & ---\\
050528 & no & 10.8 & 4.40 & --- & 849.0 & --- & --- & 2.5 & 17.2 & 3.88 & 17.96 & 1.0 & --- & 0.43 & --- & ---\\
050713B & yes & 75.0 & 45.7 & 90.2 & 2.27 & 2.88 & 0.7 & 3.3 & 18.2 & 3.80 & 19.32 & 0.5 & --- & 1.25 & $3.08\pm0.07$ & 52.947\\
050716 & yes & 69.0 & 63.2 & 70.2 & 3.83 & 1.68 & 0.32 & 3.8 & 19.8 & 4.29 & 20.61 & 8.5 & IR & 0.29 & $1.50\pm0.05$ & 58.970\\
050925 & no & 0.068 & 0.75 & --- & 1.66 & --- & --- & 3.3 & 19.0 & 3.69 & 21.12 & 2.0 & --- & 5.69 & --- & ---\\
051211A & no & 4.2 & 9.2 & --- & --- & --- & --- & 353.4 & 20.86 & 413.4 & 21.72 & 60.5 & --- & 0.32 & --- & ---\\
051211B & yes & 80.0 & 20.0 & 0.13 & 179.7 & 1.16 & 1.0 & 66.1 & 16.5 & 67.4 & 17.0 & 1.0 & --- & 1.26 & $0.82\pm0.06$ & 0.322\\
060114 & no & 100.0 & 13.0 & --- & --- & --- & --- & 2.2 & 19.0 & 37.59 & 20.70 & 15.5 & --- & 0.09 &--- & ---\\
060116 & yes & 113.0 & 26.0 & 0.9 & 2.57 & 0.95 & 1.1 & 18.72 & 18.58 & 39.98 & 20.0 & 8.5 & IR & 0.69 & $1.01\pm0.04$ & 0.990\\
060121 & yes & 2.0 & 43.0 & 0.46 & 176.70 & 1.08 & 1.07 & 50.23 & 19.5 & 175.56 & 22.22 & 20.83 & O & 0.04 & $1.20\pm0.04$ & 0.644\\
060204C & no & 60.0 & 3.5 & 0.001 & 2.6 & --- & --- & 6.42 & 18.7 & 6.89 & 19.39 & 1.0 & --- & 0.49 & --- & ---\\
060319 & yes & 12.0 & 2.7 & 2.2 & 2.88 & 1.02 & 1.10 & 7.0 & 19.0 & 9.90 & 21.63 & 3.0 & IR & 0.06 & $0.95\pm0.02$ & 4.559\\
060602A & yes & 60.0 & 16.0 & --- & --- & --- & --- & 7.72 & 15.0 & 36.08 & 16.83 & 21.0 & O & 0.07 & $1.15\pm0.13$ & 1.595\\ 
060602B & yes & 9.0 & 1.8 & 0.32 & 1.38 & 1.05 & 2.1 & 19.2 & 18.0 & 38.20 & 20.85 & 2.5 & --- & 95.63 & --- & ---\\
060825 & yes & 8.1 & 9.8 & 3.57 & 1.1 & 0.87 & 0.64 & 4.43 & 18.7 & 7.31 & 19.47 & 3.0 & --- & 1.55 & $0.96\pm0.05$ & 1.626\\
060901 & yes & 20.0 & 7.0 & 0.26 & 226.0 & 1.7 & 1.1 & 142.8 & 21.0 & 177.6 & 22.10 & 20.0 & --- & 1.85 & $1.38\pm0.08$ & 0.553\\
060923A & yes & 51.7 & 8.7 & 4.9 & 1.35 & 2.7 & 1.1 & 2.8 & 19.0 & 8.88 & 19.90 & 2.0 & IR & 0.16 & $1.69\pm015$ & 1.732\\
060923C & yes & 76.0 & 16.0 & 85.0 & 3.38 & 3.4 & 0.85 & 4.22 & 19.0 & 14.50 & 20.3 & 3.67 & IR & 0.17 & $3.09\pm0.09$ & 0.700\\
060929 & yes & 12.4 & 2.8 & 0.53 & 1.53 & 0.79 & 1.3 & 21.13 & 19.0 & 25.88 & 20.36 & 3.0 & --- & 0.13 & $1.07\pm0.04$ & 0.396\\
060930 & no & 20.0 & 2.5 & --- & --- & --- & --- & 1.98 & 19.5 & 6.45 & 20.85 & 1.5 & --- & 0.22  & --- & ---\\
061006 & yes & 0.5 & 14.3 & 0.19 & 2.38 & 2.26 & 0.7 & 22.61 & 18.0 & 23.05 & 18.20 & 0.5 & O & 0.85 & $0.77\pm0.04$ & 0.306\\
061210 & yes & 0.2 & 11.0 & --- & --- & --- & --- & 2.39 & 17.0 & 2.8 & 18.20 & 0.5 & --- & 0.09 & $2.20\pm0.42$ & 118664.2\\ 
070103 & yes & 19.0 & 3.4 & 0.38 & 1.15 & 1.4 & 1.3 & 23.7 & 19.0 & 25.20 & 19.45 & 1.0 & --- & 0.18 & $1.42\pm0.05$ & 3.00 \\
070219 & yes & 17.0 & 3.2 & 0.12 & 1.37 & 2.2 & 1.0 & 51.4 & 19.9 & 59.09 & 20.64 & 5.0 & --- & 0.09 & $0.75\pm0.28$ & 0.104 \\
070220 & yes & 129.0 & 106.0 & 16.5 & 1.32 & 1.76 & 0.55 & 1.93 & 19.5 & 34.01 & 20.47 & 10.0 & --- & 2.41 & $0.76\pm0.09$ & 3.834 \\
070223 & yes & 89.0 & 17.0 & 92.0 & 1.83 & 2.3 & 0.7 & 18.7 & 21.4 & 38.84 & 22.29 & 13.0 & IR & 0.04 & $0.88\pm0.06$ & 0.407 \\
070412 & yes & 34.0 & 4.8 & 33.0 & 1.02 & 0.98 & 1.2 & 14.1 & 21.0 & 35.68 & 21.89 & 13.0 & --- & 0.06 & $1.02\pm0.03$ & 1.34 \\
070521 & yes & 37.9 & 80.0 & 3.2 & 1.28 & 0.5 & 1.11 & 2.35 & 19.3 & 31.77 & 22.70 & 12.0 & --- & 0.07 & $0.36\pm0.16$ & 8.02 \\
070531 & yes & 44.0 & 11.0 & --- & 2.13 & --- & --- & 11.9 & 18.2 & 14.64 & 19.45 & 3.0 & --- & 1.00 & $1.32\pm0.17$ & 1.01 \\
070704 & yes & 380.0 & 59.0 & 285.0 & 2.55 & 0.87 & 0.85 & 239.5 & 21.1 & 254.34 & 22.11 & 30.0 & --- & 5.01 & $0.92\pm0.10$ & 0.150 \\
070721A & yes & 3.4 & 0.71 & 0.823 & 1.43 & 2.97 & 1.24 & 229.2 & 19.0 & 261.72 & 20.08 & 15.5 & --- & 0.04 & $0.78\pm0.04$ & 0.089 \\
070721B & yes & 340.0 & 36.0 & 24.5 & 1.53 & 0.9 & 0.48 & 327.0 & 18.5 & 364.44 & 19.31 & 15.17 & --- & 0.08 & $1.63\pm0.07$ & 0.195 \\
070808 & yes & 32.0 & 12.0 & 1.0 & 1.9 & 3.5 & 1.8 & 2.35 & 19.7 & 36.69 & 20.42 & 16.0 & --- & 0.06 & $0.93\pm0.03$ & 0.283 \\
070810B & yes & 80.0 & 0.12 & --- & 1.03 & --- & --- & 2.80 & 20.0 & 40.63 & 21.15 & 14.0 & --- & 0.14 & --- & --- \\

\enddata

\tablecomments{Notes - Reference for Duration, $\it f$, F$_{\rm X}$, $\Delta~t_{\rm X}$, $\alpha_{\rm X}$ and $\beta_{\rm X}$ : ${\it http://heasarc.gsfc.nasa.gov/docs/swift/archive/grb\_table/}$. References for OT : GRB~050124 : \cite{ber1,ber2}; GRB~050716 : \cite{tan1,rol2}; GRB~060116 : \cite{koc1,koc2,swan1,male0,tan2}; GRB060121: \cite{lev0,male1,hear1,hear2}; GRB~060319 : \cite{tan2b}; GRB~060602A : \cite{jen1}; GRB~060923A : \cite{tan3,fox1,fox0}; GRB~060923C: \cite{cov1,dav2}; GRB~061006 : \cite{male2,male3}; GRB~070223 : \cite{castro,rol}. Reference for $A_{R}$ : the values of $A_{R}$ come from the NED extinction calculator and are calculated from the list of $A_{\lambda}/E(B-V)$ reported in Table 6 of Schlegel et al. (1998) assuming an average value $R_{V}= A_{V}/E(B-V)= 3.1$.}

\end{deluxetable}
\end{landscape}


\clearpage

\setcounter{table}{2}

\begin{deluxetable}{@{}lccc}
\tabletypesize{\normalsize}
\tablewidth{0pt}
 \tablecaption{Temporal decay index $\alpha$ and spectral index
 $\beta$ in the slow cooling regime as functions of the electron
 spectral index $p$ for ISM ($\rho$=constant) or wind-like ambient
 medium ($\rho=R^{-2}$). The cases of energy injection ($L \propto
 t^{-q}$) and no energy injection (q=1) are considered (e.g. Zhang et
 al. 2006). } \label{tab3}
\tablehead{\colhead{Slow cooling} & \colhead{$\beta$} & \colhead{$\alpha$ (no injection)} & \colhead{$\alpha$ (injection)}}
\startdata
$\nu_{\rm m} < \nu < \nu_{\rm c}$ (ISM) & $\frac{p-1}{2}$ & $\frac{3(p-1)}{4}$ & $\frac{(2p-6)+(p+3)q}{4}$\\
$\nu_{\rm m} < \nu < \nu_{\rm c}$ (wind) & $\frac{p-1}{2}$ & $\frac{3p-1}{4}$ & $\frac{(2p-2)+(p+1)q}{4}$\\
$\nu_{\rm c} < \nu$ (ISM/wind) & $\frac{p}{2}$ & $\frac{3p-2}{4}$ & $\frac{(2p-4)+(p+2)q}{4}$\\
\enddata
\end{deluxetable}


\clearpage

\setcounter{table}{3}

\begin{deluxetable}{@{}cccc}
\tabletypesize{\normalsize}
\tablewidth{0pt}
 \tablecaption{Mean values and standard deviations for the
 distribution of the luminosity rest-frame L$_{R}$ at different
 times. LZ06 refer to values from Liang $\&$ Zhang (2006) and N06 to
 values from Nardini et al. (2006). See text for more details.}
\label{tablum}
\tablehead{\colhead{} & \colhead{} & \colhead{log~L$_{R}$} & \colhead{} \\
\colhead{Sample} & \colhead{(t = 1 day)} & \colhead{(t = 12 hours)} & \colhead{(t = 10 min)} \\
 \colhead{} & \colhead{[erg~s$^{-1}$]} & \colhead{[erg~s$^{-1}$~Hz$^{-1}$]} & \colhead{[erg~s$^{-1}$]}}
\startdata
\hline
LZ06 dim & $44.66\pm0.41$ & --- & --- \\
LZ06 lum & $46.15\pm0.77$ & --- & --- \\
\hline \hline
N06 & --- & $30.65\pm0.28$ & --- \\
\hline \hline
Our result & $44.25\pm0.70$ & $29.54\pm0.67$ & $46.55\pm1.23$ \\
\enddata
\end{deluxetable}

\end{document}